%% file: fieni-vPUE.tex
\documentclass[lettersize,journal]{IEEEtran}
\IEEEoverridecommandlockouts

\usepackage{wrapfig}
\usepackage{cite}
\usepackage{amsmath,amssymb,amsfonts}
\usepackage{algorithmic}
\usepackage{graphicx}
\usepackage{textcomp}
\usepackage{xcolor}
\usepackage{xspace}
\usepackage{hyperref}
\usepackage{multirow}
\usepackage{bookmark}
\usepackage{tabularx}
\usepackage{booktabs}
\usepackage{glossaries}

\newcommand{\SYS}[0]{$x$PUE\xspace}

\newcommand{\powerapi}[0]{\textsc{PowerAPI}\xspace}

\newacronym{dc}{DC}{\em Data Center}
\newacronym{vm}{VM}{\em Virtual Machine}
\newacronym{pue}{PUE}{\em Power Usage Effectiveness}
\newacronym{kpi}{KPI}{\em Key Performance Indicator}
\newacronym{wue}{WUE}{\em Water Usage Effectiveness}
\newacronym{spue}{\sc sPUE}{\em server\,PUE}
\newacronym{vpue}{\sc vPUE}{\em virtual\,PUE}
\newacronym{cpue}{\sc cPUE}{\em cloud\,PUE}
\newacronym{gpue}{\sc gPUE}{\em global\,PUE}
\newacronym{gwue}{\sc gWUE}{\em global\,WUE}
\newacronym{gcue}{\sc gCUE}{\em global\,CUE}
\newacronym{dcie}{\sc DCiE}{\em Data Center Infrastructure Efficiency}
\newacronym{cue}{CUE}{\em Carbon Usage Effectiveness}
\newacronym{xaas}{\sc XaaS}{\em Everything-as-a-Service}
\newacronym{maas}{\sc MaaS}{\em Metal-as-a-Service}
\newacronym{iaas}{\sc IaaS}{\em Infrastructure-as-a-Service}
\newacronym{caas}{\sc CaaS}{\em Container-as-a-Service}
\newacronym{paas}{\sc PaaS}{\em Platform-as-a-Service}
\newacronym{saas}{\sc SaaS}{\em Software-as-a-Service}
\newacronym{ipmi}{IPMI}{\em Intelligent Platform Management Interface}
\newacronym{rapl}{RAPL}{\em Running Average Power Limit}
\newacronym{hwpc}{\sc HwPC}{\em Hardware Performance Counters}
\newacronym{cdf}{CDF}{\em Cumulative Distribution Function}
\newacronym{tor}{\sc ToR}{\em Top of Rack}
\newacronym{ppue}{$p$PUE}{\em partial~PUE}

\def\BibTeX{{\rm B\kern-.05em{\sc i\kern-.025em b}\kern-.08em
    T\kern-.1667em\lower.7ex\hbox{E}\kern-.125emX}}

\usepackage{orcidlink}

\begin{document}

\title{$x$PUE: Extending Power~Usage~Effectiveness Metrics for~Cloud~Infrastructures%
}

\author{
    \IEEEauthorblockN{
        Guillaume Fieni\orcidlink{0000-0002-0165-6824},\IEEEauthorrefmark{1}\,\IEEEauthorrefmark{2} \and
        Romain Rouvoy\orcidlink{0000-0003-1771-8791},\IEEEauthorrefmark{2}\,\IEEEauthorrefmark{1} \and
        Lionel Seinturier\orcidlink{0000-0003-0006-6088}\IEEEauthorrefmark{2}\,\IEEEauthorrefmark{1}\\
    }
    \IEEEauthorblockA{
        \IEEEauthorrefmark{1}\,{Inria, France}\\
        \IEEEauthorrefmark{2}\,{Univ.\,Lille, CRIStAL, UMR\,CNRS\,9189, France}
    }
}%

\markboth{ }
{Shell \MakeLowercase{\textit{et al.}}: A Sample Article Using IEEEtran.cls for IEEE Journals}

\maketitle

\input{sections/abstract}

\input{sections/introduction}
\input{sections/related-work}
\input{sections/contribution}

\input{sections/implementation}

\input{sections/validation}

\input{sections/recommendations}
\input{sections/conclusion}

\bibliographystyle{plainurl}
\bibliography{IEEEabrv,references}
\end{document}

%% file: sections/abstract.tex
\begin{abstract}
    The energy consumption analysis and optimization of data centers have been an increasingly popular topic over the past few years.
    It is widely recognized that several effective metrics exist to capture the efficiency of hardware and/or software hosted in these infrastructures.
    Unfortunately, choosing the corresponding metrics for specific infrastructure and assessing its efficiency over time is still considered an open problem.
    For this purpose, energy efficiency metrics, such as the \acrfull{pue}, assess the efficiency of the computing equipment of the infrastructure.
    However, this metric stops at the power supply of hosted servers and fails to offer a finer granularity to bring a deeper insight into the \acrlong{pue} of hardware and software running in cloud infrastructure.
    
    Therefore, we propose to leverage complementary \acrshort{pue} metrics, coined \SYS, to compute the energy efficiency of the computing continuum from hardware components, up to the running software layers.
    Our contribution aims to deliver real-time energy efficiency metrics from different perspectives for cloud infrastructure, hence helping cloud ecosystems---from cloud providers to their customers---to experiment and optimize the energy usage of cloud infrastructures at large.
\end{abstract}

%% file: sections/introduction.tex
\section{Introduction}\label{sec:introduction}
Modern \glspl{dc} are continuously trying to maximize the \gls{pue} of their infrastructure to reduce their operating cost, and eventually their carbon emissions~\cite{ISO:30134-2}.
In the context of cloud providers, \gls{pue} is increasingly adopted and communicated as a \gls{kpi} reflecting the energy efficiency of the delivered solution.
While the optimal \gls{pue} is $1.0$, most infrastructures in 2023 stagnate at a \gls{pue} of $1.58$ on average, according to a survey from the Uptime Institute~\cite{uptime-insitute:industry-pue-2023}.
For years, reducing the \gls{pue} of \gls{dc} has become an active research area where actors compete to propose the most efficient cooling techniques, consider renewable energies as part of their electricity mix, and improve the utilization of IT resources.

However, the emergence of the cloud computing paradigm has strengthened the role and the impact of software layers on the energy consumption of these \glspl{dc}.
While the \gls{spue} has only been mentioned in the literature~\cite{barroso2013datacenter}, its scope remains limited to hardware layers, hence failing to address the software layers spectrum.
In particular, we believe that both virtualization technologies and control planes may shuffle reported energy efficiency indicators, if not appropriately optimized.
One could, for example, envision scenarios where the \gls{pue} of a \gls{dc} is reported as optimal, but the combined inefficiency of the software layers deployed in the hosted servers does not reflect the savings achieved by the building infrastructure, hence misleading cloud customers.
This is all the more prevalent as cloud provider are intensively communicating on \gls{pue} as a marketing argument to attract customers.\footnote{\url{https://www.google.com/intl/en/about/datacenters/efficiency/}}$^,$\footnote{\url{https://datacenters.microsoft.com/sustainability/efficiency/}}

We, therefore, propose to push the limits of the state-of-the-art hardware metrics---namely, the \gls{pue} and \gls{spue}---and to introduce a novel software metric---coined \gls{vpue}---to better reflect the end-to-end energy consumption of cloud computing operators by reporting on the \gls{pue} of software layers, which can be indefinitely stacked.
More specifically, we introduce \SYS{} as a family of composable metrics that can be used to study and assess the energy efficiency of cloud infrastructures across the hardware-software continuum.

Using \SYS{}, the cloud ecosystem---including cloud providers and their customers---can investigate energy wastes in the software layers of their infrastructure and take appropriate actions to reduce them.
For example, our experiments highlight the impact of platform control planes as well as the importance of operating large-scale infrastructure at their maximum capacity.

To summarize, the four contributions reported in this article are:
\begin{enumerate}
    \item A formalisation of the \gls{spue} metric, which has been mentioned by~\cite{barroso2013datacenter}, but never properly defined,
    \item The introduction of \SYS{} as a family of new and complementary \gls{pue} metrics to cover all the software layers of cloud infrastructures,
    \item The implementation of \SYS{} metrics using the \powerapi{} framework~\cite{colmant:2015},
    \item The study of \SYS{} metrics across multiple cloud deployments to deliver insightful feedback on the key factors impacting the \acrlong{pue} of cloud infrastructures.
\end{enumerate}

This article starts by delivering background on state-of-the-art \gls{pue} metrics and their limitations (cf. Section~\ref{sec:related-work}) before introducing our contribution (cf. Section~\ref{sec:contribution}).
Then, we detail the implementation of \SYS{} as a real-time energy effectiveness metric for cloud infrastructures (cf. Section~\ref{sec:implementation}) and we assess its validity across multiple cloud deployments (cf. Section~\ref{sec:validation}).
We conclude in Section~\ref{sec:conclusion}.

%% file: sections/related-work.tex
\section{Related Work}\label{sec:related-work}
While the literature in our area of \gls{dc} effectiveness is abundant, we choose to focus on the metrics of growing interest in the cloud industry: the \gls{pue} and its variants.

\paragraph{\acrfull{pue}}
The standard \textsf{ISO/IEC 30134-2:2016}~\cite{ISO:30134-2} defines the \gls{pue} as a metric reflecting the energy efficiency of a \gls{dc}.
More specifically, the \gls{pue} is defined as the ratio of the overall energy consumed at the \gls{dc} level to the energy consumed by hosted IT equipment, thus estimated as:
\begin{equation}\label{eq:pue}\small
    \text{PUE} = \frac{\sum{Energy(\textsf{DC})}}{\sum{Energy(\textsf{IT})}} = 1 + \frac{\sum{Energy(\textsf{non-IT})}}{\sum{Energy(\textsf{IT})}}
\end{equation}
where $\sum{Energy(\textsf{DC})}$ is the sum of energies drawn by what is not considered as a computing device, but is required to operate a \gls{dc} (so-called \textsf{non-IT}), such as lighting, air conditioning, etc., and the \textsf{IT} equipment.

The \gls{pue} is a widely-adopted metric, often cited by major cloud providers to demonstrate their progress in \gls{dc} efficiency.\footnote{\href{https://azure.microsoft.com/en-us/blog/how-microsoft-measures-datacenter-water-and-energy-use-to-improve-azure-cloud-sustainability/}{How Microsoft measures datacenter water and energy use to improve Azure Cloud sustainability}}
While the ideal \gls{pue} is $1.0$, the average \gls{pue} for a \gls{dc} in 2020 was $1.58$~\cite{uptime-insitute:industry-pue-2023}.
For example, Google publishes quarterly and trailing 12-month \gls{pue} values going back to 2008 for their \gls{dc} hosted globally and reports a fleet-wide \gls{pue} of $1.10$ for 2021~\cite{google:dc-effiency}.

Nonetheless, \gls{pue} has been criticized when adopted as a measure of efficiency because it only considers energy and does not consider the effective usage of the computational resources~\cite{brady:2013,dayarathna:2016}.
This means that a \gls{pue} can mechanically decrease by artificially increasing the IT workload, hence increasing the energy consumption of servers to reduce the impact of \textsf{non-IT} equipments.

\paragraph{\gls{dcie}}
The \gls{dcie} is the reciprocal of the \gls{pue}, defined as:
\begin{equation}\small
    \text{DCiE} = \frac{{1}}{{\text{PUE}}} = \frac{\sum{Energy(\textsf{IT})}}{\sum{Energy(\textsf{DC})}}
\end{equation}

One can observe that, although \gls{pue} and \gls{dcie} are the most commonly-used metrics to compare the energy efficiency of \gls{dc}, they only assess the global energy efficiency and fail to provide any insight into the IT efficiency in particular~\cite{schaeppi:2012}.

Furthermore, the scopes of \gls{pue} and \gls{dcie} do only capture the building efficiency of a \gls{dc}, hence stopping at the power outlet of hosted computing units.
This partial coverage is particularly critical in the context of cloud computing, which does not limit itself to building infrastructures, but heavily rely on computing resources (servers, routers, etc.) and software platforms to operate a wide diversity of \gls{xaas} offers---ranging from \gls{iaas} to \gls{saas}.
By focusing and communicating on \gls{pue}, cloud providers might hinder the overhead of these computing resources and software platforms, hence misleading stakeholders about the end-to-end efficiency of their products.

\paragraph{\gls{cue}}
The \gls{cue} (in kilograms of carbon dioxide per kilowatt-hours: kgCO2eq) aims to assess the efficiency of the energy used for the \gls{dc}:
\begin{equation}\small
    \text{CUE} = \frac{\sum{Emission_{CO2}(\textsf{DC})}}{\sum{Energy(\textsf{DC})}}
\end{equation}
The \gls{cue} does not take into account the embodied emissions accountable to the manufacturing of the \gls{dc} or its equipment.

However, the \gls{cue} includes the carbon emissions due to the mix of energy being used by the \gls{dc} in production.

\paragraph{\gls{spue}}
Introduced by Barroso~\emph{et~al.} in 2013~\cite{barroso2013datacenter}, the \gls{spue} is computed as the ratio of the server input power to its useful component power, including all the parts directly involved in the computations, namely motherboard, disks, CPUs, DRAM, GPU, I/O, etc.
\gls{spue} aims to quantify the efficiency of individual servers and the authors report that state-of-the-art \gls{spue} should be less than $1.2$ at the time of writing their book.
In particular, low \gls{spue} should result from efforts in delivering optimized supply and cooling steps.

However, to the best of our knowledge, neither the cloud operators nor the literature adopted this indicator to report on best practices in the design of hardware servers and, eventually, cloud computing offers.

\paragraph{DWPE}
\emph{Data center Workload Power Efficiency} (DWPE) is another metrics introduced by Wilde~\emph{et~al.} in 2019~\cite{DBLP:conf/hpcs/WildeAPSHBLC14} to assess the energy efficiency of the \emph{High Performance Computing} (HPC) infrastructures.
While DWPE complements \gls{spue} by considering application workloads, it fails to appropriately evaluate the various software layers of cloud infrastructures by focusing on performance objectives of HPC applications.

\subsubsection*{Limitations \& Opportunities}\label{sec:related-work:limitations-opportunities}
Most energy efficiency indicators aim to evaluate global---at a \gls{dc} granularity---or specific hardware parts---at a baremetal granularity---of the cloud infrastructures.
However, none of them proposes an end-to-end approach from the scale of \gls{dc} to individual servers and even the hosted software services, no matter their location and potentially distributed nature.
In particular, in the context of cloud infrastructures, providing an end-to-end energy indicator should allow deeper analysis and tuning of the infrastructure hardware and software components to optimize energy usage at large.

In this article, we therefore propose complementary metrics to analyze, in-depth, the \gls{pue} of cloud infrastructures, specifically the virtual layers introduced by these infrastructures.
Such a family of complementary \gls{pue} metrics are intended to quantify, for 1\,Watt consumed by a cloud native service, how many Watts are effectively required by the hosting \gls{dc}.

%% file: sections/contribution.tex
\section{\SYS{}: A New Family of \gls{pue} Metrics}\label{sec:contribution}
\SYS{} groups a family of \gls{pue}-related metrics that can be easily adopted by cloud providers, and more generally service providers, to estimate the \gls{pue} of their infrastructure in the deep, including the software platforms they may operate.
\SYS{} extends standard \gls{pue} and \gls{spue} metrics by delivering insights beyond the power supply of physical servers.

\subsection{Overview of \SYS{}}
\begin{figure}
    \centering
	\includegraphics[width=.9\linewidth]{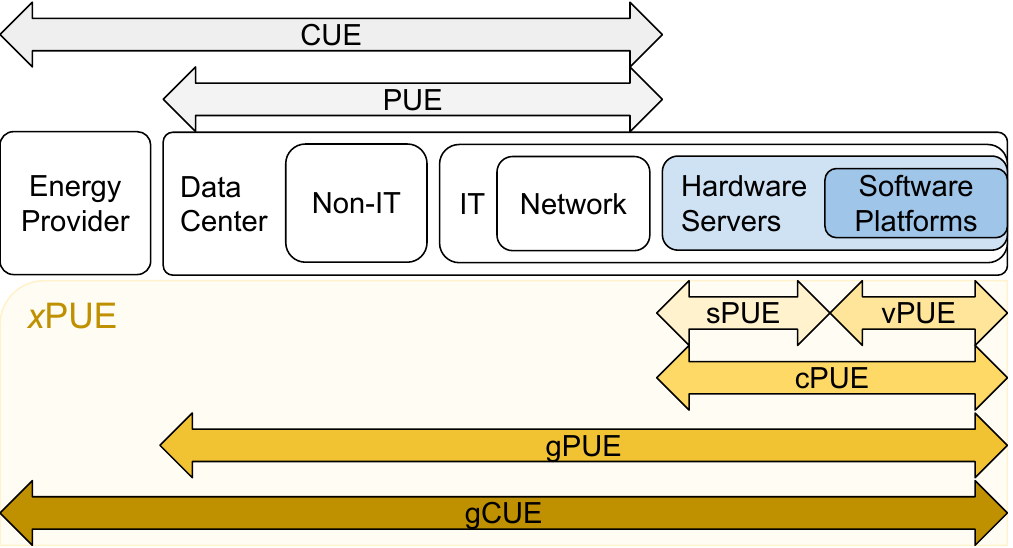}
	\caption{Efficiency coverage of state-of-the-art and \SYS{} metrics (highlighted in grey and yellow, respectively).}
	\label{fig:contribution:xpue}
\end{figure}

In this article, we specifically leverage the \gls{spue} mentioned---but never formalized---by Barroso~\emph{et~al.}~\cite{barroso2013datacenter} to introduce the \gls{vpue} metrics and highlight resource usage effectiveness at the scales of hardware and virtual layers, respectively.
Figure~\ref{fig:contribution:xpue} depicts the complementarity of \SYS{} metrics with state-of-the-art metrics, including \gls{pue}, \gls{spue}, and \gls{cue}.
One can observe that, while \gls{pue} and \gls{cue} are delivering an in-breadth coverage of the resource usage effectiveness of \gls{dc} buildings along different perspectives---energy and carbon, respectively---\gls{spue} and \gls{vpue} brings more in-depth insights by investigating how these resources are consumed within hardware clusters and servers and, ultimately, cloud software services and platforms.
As the energy consumption of the hardware servers and software services is proportional to their usage, they can exhibit significant variability when it comes to energy efficiency.
In the context of this article, we are therefore interested in investigating the end-to-end resource usage effectiveness of cloud services, by covering hardware and software layers.
By doing so, we intend to expose the real effectiveness of cloud infrastructures and to contribute to a more transparent exposure of how much energy is consumed by cloud providers for each functional unit, whenever operating a \gls{maas}, \gls{iaas}, or a \gls{caas} offer.
By focusing on the software layers that are not covered by existing metrics, we intend to raise new environmental challenges for cloud infrastructures, thus encouraging the ecosystem to maximize the end-to-end power usage effectiveness of cloud services, hence going beyond the sole effectiveness of \gls{dc} buildings, as currently covered by the \gls{pue}.
This more holistic analysis of the energy and carbon efficiency of cloud infrastructures is brought by the definition of \gls{gpue} and \gls{gcue} indicators, which are introduced in the following sections.

\subsection{\gls{spue}: Assessing Cloud Servers \gls{pue}}\label{sec:hpue}
The \gls{spue} aims to estimate the \acrlong{pue} at the scale of a server~\cite{barroso2013datacenter}.
This metric is particularly relevant in the cloud when operating a \gls{maas} (also known as \emph{bare-metal}) offer, which consists in delivering a hardware server to the customer.
In this context, the cloud provider is in charge of hosting and eventually assembling the delivered server.
While the \gls{pue} stops at the power outlet of hosted servers, the \gls{spue} metric intends to dive into the integrated components to capture the overhead imposed by the cooling systems and power supplies inside the server, as mentioned by~\cite{barroso2013datacenter}.
In this context, we formalize the \gls{spue} as the ratio between the energy consumed by the \textsf{IT} equipment (cf. Equation~\ref{eq:pue}) and the energy directly consumed by the \textsf{hardware} components contributing the computations of the cloud infrastructure (incl. CPU, GPU, memory, disk, controllers, etc.), as follows:
\begin{equation}\label{eq:hpue}\small
    \gls{spue} = \frac{\sum{Energy(\textsf{IT})}}{\sum{Energy(\textsf{hardware})}}
\end{equation}

This metric, therefore, indicates how much energy is consumed by the physical server for each unit of computation delivered by the CPU and other critical components.
By formalizing the \gls{spue}, we are interested in highlighting the efforts spent by cloud providers to deliver energy-proportional bare-metal solutions---\emph{i.e.}, limiting the cost of cooling components and optimizing the supply of direct current.
We believe that this additional dimension is important to capture, as there is no guarantee that the \gls{pue} reflects this overhead.\footnote{The \gls{pue} might eventually reflect, indirect, side-effects of server emissions like a reduced consumption of the air cooling system of a \gls{dc}.}
While the \gls{pue} is usually computed at the scale of a \gls{dc}, the \gls{spue} can be computed at the scale of a single server, a rack, a cluster, a room or a \gls{dc}, depending on the scope of the analysis.
Server-scale \gls{spue} can be used to improve the design of individual servers to reduce potential energy waste, which \gls{dc}-scale \gls{spue} is intended to report on the average efficiency of a farm of, potentially heterogeneous, production servers.
Therefore, beyond the reported numbers, we strongly recommend indicating the scale at which the \gls{spue} has been analyzed.
Additionally, we recommend listing the \textsf{hardware} components that have been considered in the computation of the \gls{spue}, as the \gls{spue} can be computed for different hardware configurations.

\gls{spue} covers general-purpose computations that can be delivered to cloud customers, but also exploited internally to operate a cloud platform.
However, no matter their nature---being \gls{iaas}, \gls{caas}, \gls{paas}, etc.---these cloud platforms are hosting a large number of software services, which are in charge of managing the cloud infrastructure and delivering the hosted services to the customers.
These software services are generally implemented as \gls{vm} or containers, and they are often deployed on top of a hypervisor or a container engine.
In this context, we are interested in investigating how much energy is consumed by the cloud software services for each unit of computation delivered to the hosted software, no matter their nature---\emph{i.e.}, \gls{vm} or containers.

\subsection{\gls{vpue}: Assessing Cloud Services \gls{pue}}\label{sec:vpue}
The metric \gls{vpue} dives into the software layers of cloud infrastructures by investigating the cost of operating large and complex software platforms, such as \href{https://www.openstack.org}{\textsc{OpenStack}} for a \gls{iaas} or \href{https://kubernetes.io}{\textsc{Kubernetes}} for a \gls{caas}.
These cloud solutions generally share two key concepts: \emph{virtualization techniques} to control access to the computational resources, and \emph{control planes} to deploy and manage the hosted services.
To capture this metric, we define the \gls{vpue} as follows:
\begin{equation}\label{eq:vpue}\small
    \gls{vpue} = \frac{\sum{Energy(\textsf{hardware})}}{\sum{Energy(\textsf{software})}}
\end{equation}
where \textsf{hardware} (taken from Equation~\ref{eq:hpue}) refers to all the server components required to operate the \textsf{software} hosted by the considered cloud platform.
The list of included \textsf{software} covers all the processes running on the hosting server, which encompass the operating system, the cloud platform, as well as other daemon processes that can be deployed by the cloud operator.
At first, the resulting ratio is intended to measure the overhead imposed on the execution of general-purpose software processes.
From this general-purpose indicator, we can leverage the principle of \gls{ppue}~\cite{ISO:30134-2} to introduce partial \gls{spue} indicators that capture the overhead imposed by different aspects of the cloud infrastructure.
This can include the overhead of hypervisors and the associated control planes that we consider as a specific \emph{scope} ($S$), which is used to filter out the software process of interest together with the hardware components they are running on:
\begin{equation}\label{eq:pvpue}\small
    \gls{vpue}(S) = \frac{\sum_{s \in S}{Energy(\textsf{hardware}_s)}}{\sum_{s \in S}{Energy(\textsf{software}_s)}}
\end{equation}
where $s \in S$ selects all the hardware and software parts that are related to the scope $S$.
As a result, \gls{vpue}(\textsc{Kubernetes}) is intended to capture the overhead of the \textsc{Kubernetes} cloud platform, while \gls{vpue}(\textsc{OpenStack}) is intended to capture the overhead of the \textsc{OpenStack}.
This scope can also be further refined to capture the overhead of dedicated planes of a cloud platform, such as comparing the efficiency of the control plane compared to data plane,  as illustrated in Section~\ref{sec:validation}.

One should also keep in mind that, unlike hardware layers, software layers can be stacked by cloud providers and/or their customers.
For example, the deployment of a \textsc{Kubernetes} cluster atop \gls{vm} hosted by a \gls{iaas} infrastructure is commonly adopted by practitioners to offer more flexibility when it comes to adjusting resource usage.

\subsection{\acrshort{cpue}: Applying \SYS Metrics to Cloud Infrastructures}\label{sec:cpue}
To account for nested virtualization practices and to better reflect the end-to-end efficiency of a cloud platform, we introduce the \gls{cpue} metric, which is a  metric computed from \gls{spue} and \gls{vpue}.
Concretely, to report this compound metric, one needs to compute the product of \SYS{} metrics, depending on the spectrum of the deployed infrastructure, as follows:
\begin{equation}\small
    \gls{cpue} = \prod_{x \in L} x\text{PUE}\text{ with $L$ the selected cloud layers}
\end{equation}
where the list of selected layers, $L$, depends on the context and the owner of the cloud infrastructure.
For example, a \gls{iaas} provider operating an \textsc{OpenStack} cloud platform will compute $\gls{cpue} = \gls{spue} \times \gls{vpue}(\textsf{OpenStack})$.
Similarly, a \textsc{Kubernetes} platform hosted on-premise will rather be reported as $\gls{cpue} = \gls{spue} \times \gls{vpue}(\textsf{Kubernetes})$.
Finally, because of the recursive property of virtualization techniques, one can imagine computing $\gls{cpue} = \gls{spue} \times \gls{vpue}(\textsf{OpenStack}) \times \gls{vpue}(\textsf{Kubernetes})$ to capture the end-to-end \gls{pue} a multi-layers \gls{caas} or \gls{paas} platform leveraging several cloud technologies.
Interestingly, stakeholders of a cloud ecosystem do not need full control of the infrastructure to estimate the \gls{cpue} of their solution.
For example, any customer of a \gls{iaas} provider can compute the \gls{cpue} of their hosted software by multiplying the \gls{vpue} of their \gls{vm} by the \gls{cpue} reported by their cloud provider.

Furthermore, beyond being a product of \SYS{} metrics, any \gls{cpue} metric can also be multiplied by the \gls{pue} of the \gls{dc} hosting the software services as follows:
\begin{equation}\label{eq:gpue}\small
    \gls{gpue} = \gls{cpue} \times \gls{pue}
\end{equation}
which is reported as the \gls{gpue} revealing, for each unit of computation performed by cloud software, how much energy is effectively consumed by the whole \gls{dc} hosting this software.
Given that \SYS{} metrics share the same properties as the \gls{pue}---\emph{i.e.}, the ideal value is $1.0$---one can observe that any waste of resource in any of the covered layers may severely impact the \acrlong{gpue} of the infrastructure, thus challenging the cloud ecosystem to pay attention to the effectiveness of their hardware and software solutions.

At this point, one can suspect that the \gls{pue} reported by the cloud industry does not reflect the real cost of the efficiency of infrastructures by only focusing on the optimizations operated on the hosting facilities and omitting the cost of the software running on top of these facilities.
We, therefore, recommend adopting the \gls{gpue} metric to share more transparently the efficiency of individual cloud offers.

\subsection{Revisiting State-of-the-Art Metrics with \gls{cpue}}\label{sec:gmetrics}
Beyond the \gls{gpue}, one can also revisit the state-of-the-art metrics to consider their global impact, beyond the \gls{dc} building optimizations.
In particular, the \gls{gcue} metric can be more accurately reported by cloud infrastructures as follows:
\begin{equation}\small
    \gls{gcue} = \gls{cpue} \times \gls{cue}
\end{equation}
where the \gls{cue} metric is extended with the cost of the cloud platform, hence better reflecting the carbon footprint of each unit of computation operated by a cloud platform.

 \SYS{} can be extended to other environmental metrics, such as \gls{dcie}, \emph{Green Power Usage Effectiveness} (GPUE), \emph{Green Energy Coefficient} (GEC), \emph{Energy Reuse Factor} (ERF) and \emph{Water Usage Effectiveness} (WUE) to cover a larger scope of the environmental impact of cloud infrastructures~\cite{priya2013survey}.

%% file: sections/implementation.tex
\section{Implementation Details of \SYS{} Metrics}\label{sec:implementation}
The implementation of the \SYS{} metrics requires reporting power measurements at a finer granularity than a power outlet.
While any software-defined power meter, like \textsc{Scaphandre}\footnote{\url{https://github.com/hubblo-org/scaphandre}} or \textsc{Kepler},\footnote{\url{https://sustainable-computing.io}} can be used to implement the \SYS metrics we propose, we leverage the \powerapi toolkit~\cite{colmant:2015,DBLP:journals/jossw/FieniARR24} to implement the runtime support \SYS{}.

\subsection{Introducing the \powerapi{} Toolkit}\label{sec:powerapi}
\powerapi is a middleware toolkit for building software-defined power meters~\cite{colmant:2015,DBLP:journals/jossw/FieniARR24}. 
Software-defined power meters are configurable software libraries that can estimate the power consumption at the scale of individual software processes in real-time.
\powerapi supports the acquisition of raw metrics from a wide diversity of sensors (\emph{e.g.}, physical meters, processor interfaces, hardware counters, OS counters) and the delivery of power consumptions via different channels (including file system, network, web, graphical).

To implement \SYS{}, we more specifically leverage the \powerapi toolkit to collect energy metrics from multiple sources of energy measurements and to aggregate them.
For the global energy measurements required by the \gls{spue}---\emph{i.e.}, $\sum{Energy(\textsf{IT})}$ in Equation~\ref{eq:hpue}---we leverage hardware power meters plugged into the power supply and \gls{ipmi} for the global power measurements of the servers.
As \gls{ipmi} measurements have a low refresh rate and a low accuracy, we recommend using hardware power meters whenever possible.
When the server combines multiple power supplies, the power measurements of all active power supplies are aggregated.

For hardware-specific measurements (CPU, DRAM, GPU\dots) required by the \gls{spue} and \gls{vpue}---\emph{i.e.}, $\sum{Energy(\textsf{hardware})}$ in Equations~\ref{eq:hpue} and~\ref{eq:vpue}---we consider the hardware power interfaces, such as \gls{rapl} for the CPU, as it is widely available and accurate on recent CPUs~\cite{Rotem:2012,Desrochers:2016}.

Regarding software power measurements---\emph{i.e.}, $\sum{Energy(\textsf{software})}$ in Equation~\ref{eq:vpue}---we use the \textsc{SmartWatts} power meter~\cite{fieni:2020,DBLP:conf/ccgrid/FieniRS21}, which automatically infers power models from hardware measurements and disaggregates power consumption among software processes.
\textsc{SmartWatts} supports power estimation both at the granularities of \glspl{vm} and containers, as well as it succeeds in estimating CPU and DRAM power consumptions in real-time with high accuracy.
Furthermore, as \textsc{SmartWatts} offers process-scale power measurements, we can therefore easily separate the power consumption of any software or service running in the infrastructure (from control/master nodes to worker/compute nodes).
Concretely, to differentiate the hosted cloud software from the control plane services, we can select and tag pre-defined groups of software.
For example, in a \textsc{Kubernetes} cluster, some services are directly related to the \textit{infrastructure}, like the container runtime, control plane, networking, and monitoring services.
Subgroups can also be defined for the services: one can compute dedicated \gls{vpue} for network and monitoring services to further analyze their energy efficiency.

\subsection{Implementing the \SYS{} Formulas}
\begin{figure}
    \centering
	\includegraphics[width=.9\linewidth]{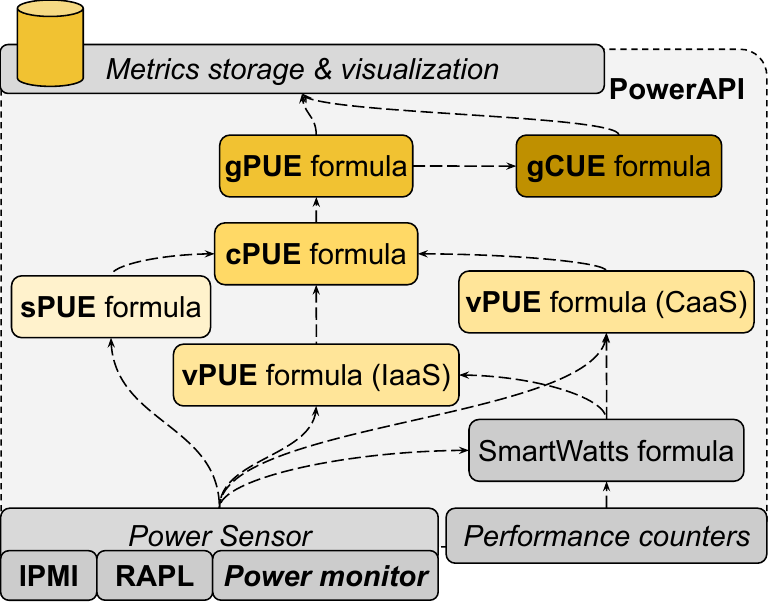}
	\caption{Deployment of \SYS{} metrics using \powerapi}
	\label{fig:implementation:deployment}
\end{figure}

We implemented each \SYS{} metric as a Python formula in the open-source library \powerapi (cf. Figure~\ref{fig:implementation:deployment}).
The \powerapi toolkit exposes a software agent that consumes input measurements from various sources (databases, message queues), processes the samples, and produces estimations through the same or another database(s).
We also leverage the \textit{de~facto} standard libraries in Python: \href{https://pandas.pydata.org}{\textsc{Pandas}} for the manipulation and analysis of samples, as well as \href{https://www.scipy.org}{\textsc{Scipy}} for the computation of the \SYS{} estimations. 

This results in the development of a family of dedicated \SYS{} formulas atop \powerapi that continuously estimate \gls{spue} and \gls{vpue} indicators and can even be further combined to report on the \gls{vpue} and \gls{gpue} compound metrics in real-time.%
As \SYS{} metrics rely on power measurements coming from multiple sources (\gls{rapl}, \gls{ipmi}, hardware, and software power meters) that are not synchronized, we resample the power measurements to a common time base.
By default, \SYS{} formula detects the shortest time window possible to resample the \textsc{Pandas} \textit{DataFrame}.
Finally, our implementation of \SYS{} leverages a database to store the power measurements received from hardware and software power meters.
These measurements are then handled by the formula components and \SYS{} metrics are then stored in the database for further analysis and visualization (\emph{e.g.}, as a \textsc{Grafana} dashboard).

\subsection{Deploying the \SYS{} Metrics}
To simplify the deployment process, \SYS{} is available as containers, which provide an environment-agnostic deployment and ease the lifecycle management of its related services.
The formula components can be deployed on any host of the cluster, or a remote server, as it only requires access to the power measurements through a message queue to work.
For example, one can use a \textsc{MongoDB} instance as a message queue to communicate the power-meters measurements through a publish-subscribe pattern and then store the \SYS{} metrics into an \href{https://www.influxdata.com/}{\textsc{InfluxDB}} \emph{Time Series DataBase} (TSDB) or a \href{https://prometheus.io/}{\textsc{Prometheus}} agent.
The \SYS{} metrics can be exposed as \href{https://grafana.com/}{\textsc{Grafana}} dashboards for environment/service-specific metrics reporting and real-time/offline analysis.
Figure~\ref{fig:implementation:deployment} depicts a typical deployment of \SYS{} to monitor the \gls{vpue} of multiple parts of a cloud infrastructure involving both \gls{iaas} (\textsc{OpenStack}) and \gls{caas} (\textsc{Kubernetes}) platforms.

%% file: sections/validation.tex
\section{Empirical Validation}\label{sec:validation}
This section builds on our implementation of \SYS{} to study the various factors that may contribute to improving or degrading the \acrlong{pue} of cloud infrastructures.
We, therefore, start by investigating the impact of \gls{spue} on different hardware configurations (cf. Section~\ref{sec:xp:hpue}), then exploring the \gls{vpue} in the context of a \gls{iaas} infrastructure (based on \textsc{OpenStack}) and a \gls{caas} infrastructure (based on \textsc{Kubernetes}).
We conclude by illustrating the \gls{cpue} in the context of a \gls{caas} deployed in a \gls{iaas}, a widely-adopted architecture in the cloud industry (cf. Section~\ref{sec:xp:cpue}).
Throughout this section, we assess the relevance of \SYS{} to evaluate the \acrlong{pue} into deeper layers of \textsc{Kubernetes} and \textsc{OpenStack} based cloud infrastructures.

\subsection{Evaluation Methodology}
We follow the experimental guidelines reported by~\cite{vanderkouwe:2018} to enforce the quality of our empirical results.
For the sake of reproducible research, \SYS{}, the necessary tools, deployment scripts, and resulting datasets are open-source and publicly available on GitHub.\footnote{Anonymized}%

\begin{table*}\small
    \centering
    \caption{Testbed hardware settings}
    \label{table:validation:hardware-settings}
	\resizebox{.9\textwidth}{!}{%
    \begin{tabular}{l|c|c|c|c|c}
    \toprule
	Provider           & \multicolumn{3}{c|}{Grid'5000}      & \multicolumn{2}{c}{\textsc{OVHcloud}} \\
    Model              & Dell PowerEdge R640    & Dell PowerEdge R7525 & Dell PowerEdge R640 &  Intel bare-metal server  & AMD bare-metal server \\
    \toprule
    \toprule
    CPU                & Intel Xeon Gold\,5220  & AMD EPYC\,7352  & Intel Xeon Gold 5218 & Intel Xeon Silver\,4214R & AMD EPYC\,7413 \\
    Generation         & Cascade Lake           & Zen\,2	      & Cascade Lake 		 & Cascade Lake             & Zen\,3 \\
    Socket(s)          & 1                      & 1 		 	  & 2 					 & 2                   	 	& 1 \\
    Cores per socket   & 18                     & 24 			  & 16 					 & 12                 		& 24 \\
    Threads per socket & 36                     & 48 			  & 32 					 & 24                  	 	& 48 \\
    Total threads per server & 36               & 48 			  & 64 					 & 48                  	 	& 48 \\
    Memory             & 96\,GB                 & 128\,GB 		  & 384\,GB 			 & 32\,GB                   & 64\,GB \\
    TDP                & 125\,W                 & 155\,W 		  & 125\,W 				 & 100\,W                   & 180\,W \\
    Cooling            & \multicolumn{3}{c|}{Air}                                        & \multicolumn{2}{c}{Water} \\
    \bottomrule
    \end{tabular}
	}
\end{table*}

\subsubsection*{Hardware Settings}
Most of our experiments are deployed on the Grid'5000 testbeds infrastructure~\cite{grid5000}, which is a large-scale and flexible testbed for experiment-driven research in all areas of computer science, with a focus on parallel and distributed computing---including cloud, HPC, Big Data, and AI.
We deploy our experimental infrastructure on $5$ nodes of the cluster \textsf{gros} located in the site of \textsf{Nancy}.
The description of the considered servers is reported in Table~\ref{table:validation:hardware-settings} (cf. 3 first columns).
This cluster is particularly interesting as each node has its power supply monitored by a hardware power meter to monitor the power consumption of the node with high accuracy and at a high frequency.

We also consider the provisioning of additional bare-metal servers from a production-scale cloud infrastructure provided by \href{https://www.ovhcloud.com/}{\textsc{OVHcloud}} (2 last columns of Table~\ref{table:validation:hardware-settings}).
We chose this cloud provider as all the hosted servers are cooled using an advanced water cooling system, which may contribute favorably to the \gls{spue}, compared to traditional air cooling.\footnote{\url{https://blog.ovhcloud.com/water-cooling-from-innovation-to-disruption-part-i/}}
Furthermore, \textsc{OVHcloud} proposes a \gls{maas} offer with a wide diversity of hardware architectures (including Intel and AMD processors) and access to server-scale power measurements.

\subsubsection*{Software Settings}
All machines of our experiment are using the \textsf{Ubuntu\,20.04\,LTS} Linux distribution with a kernel version \texttt{5.4.0-121-generic}, where only a minimal set of daemons are running in the background.

For the \textsc{Kubernetes} cluster, the deployment is done using \textsf{kubeadm} and the version deployed is the \textsf{1.21}.
The container runtime is \textsf{Containerd} version \textsf{1.6.6} and the \emph{Container Network Interface} (CNI) deployed is \href{https://github.com/flannel-io/flannel}{\textsc{Flannel}} version \textsf{0.18.1}.

For the \textsc{OpenStack} cluster, the deployment is done using \href{https://microstack.run}{\textsc{MicroStack}}, which deploys \textsc{OpenStack} version \textsf{Ussuri} through the \href{https://snapcraft.io}{\textsc{Snap}} package manager.
This allows us to quickly deploy an \textsc{OpenStack} instance in self-contained packages that support individual monitoring through Linux Cgroups. 
This choice considerably eases the deployment and the reproducibility of our results, while remaining representative of real-world deployments.

\subsubsection*{Input Workloads}
For both deployments, we use a state-of-the-art benchmark tool for Linux, \href{https://launchpad.net/stress-ng}{\textsc{stress-ng}}, to simulate various resource-intensive workloads that stress various parts of the system.
Therefore, during each experiment, containers and \gls{vm} will be started with a random resource workload that will be maintained until it is stopped.
This allows us to generate a base workload on the infrastructure and to stress multiple parts of the software services and underlying hardware under varying loads, from resource-scarce to resource-intensive CPU workloads.

Our approach does not prevent to use additional benchmarks to stress the system, like \textsc{SPEC}\footnote{\url{https://www.spec.org/benchmarks.html}} or \textsc{Phoronix},\footnote{\url{https://openbenchmarking.org/tests/pts}} to further investigate the impact of the input workload on the energy efficiency of the cloud infrastructure, as well as memory-intensive benchmarks.

While synthetic, these input workloads are intended to mimic real-world workloads that can be found in cloud infrastructures~\cite{DBLP:conf/ic2e/JacquetLR23}, and are used to stress the system to its maximum capacity, therefore exploring a wide spectrum of values for the \gls{spue} and \gls{vpue} indicators we introduced.

\subsubsection*{Power Meters}
To monitor the power consumption of the servers, we use the available hardware power meters attached to the input of the power supply of every machine in the cluster.
The measurements are automatically reported to an \textsc{InfluxDB} at a frequency $50$ measurements per second, which is more than required for our experiments.

Regarding software measurements, \textsc{SmartWatts} requires the deployment of two components, the \texttt{Sensor} which monitors the \gls{hwpc} and needs to be deployed on every node, and the \texttt{Formula} component which computes the power estimations.
In all our experiments, we configure the \texttt{Sensor} components to report on power estimations once per second ($\beta=1\,\textsc{Hz}$), and the \textsc{Formula} component with an error threshold of $\alpha=5\,W$ at the scale of the CPU package (\texttt{PKG}).%

\subsection{\gls{spue} Experiments}\label{sec:xp:hpue}
\begin{figure}
    \centering
	\includegraphics[width=\linewidth]{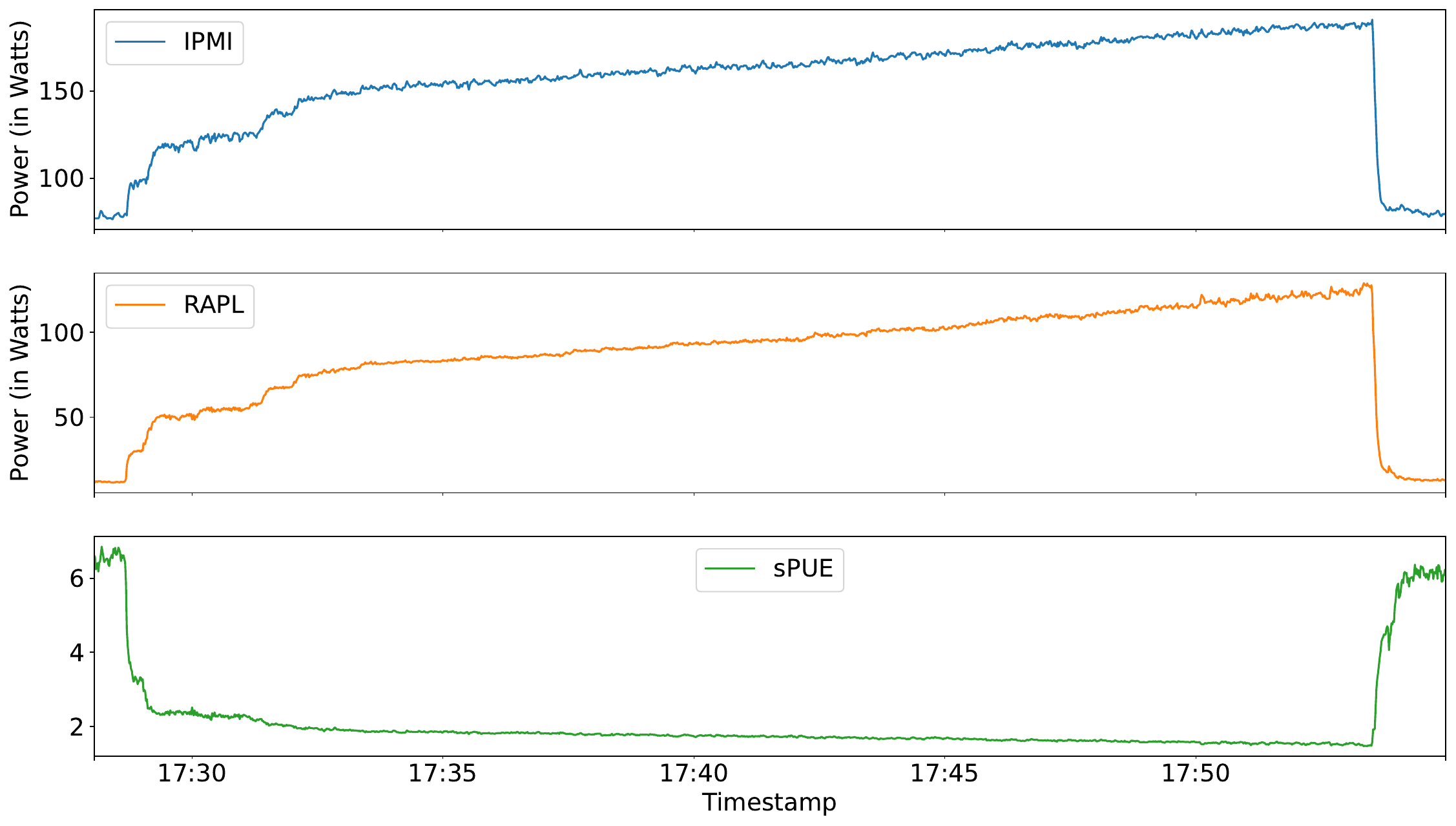}
	\caption{Evolution of the \gls{spue} over time and increasing workload}
	\label{fig:xp:k8s:hpue-time}
\end{figure}

\paragraph{Server setting.}
To evaluate the relevance of \gls{spue}, we run our benchmark designed to stress twice the maximum CPU load of the Intel Xeon Gold server.
This aims to investigate the behavior of the hardware infrastructure in a situation of resource over-commitment.
Figure~\ref{fig:xp:k8s:hpue-time} reports on the evolution of the \gls{spue} over time, when increasing CPU workload.
One can first observe that idle servers seriously degrade the energy efficiency of cloud infrastructures, with a $\gls{spue} >4$.

\begin{figure*}
    \centering
	\includegraphics[width=.7\linewidth]{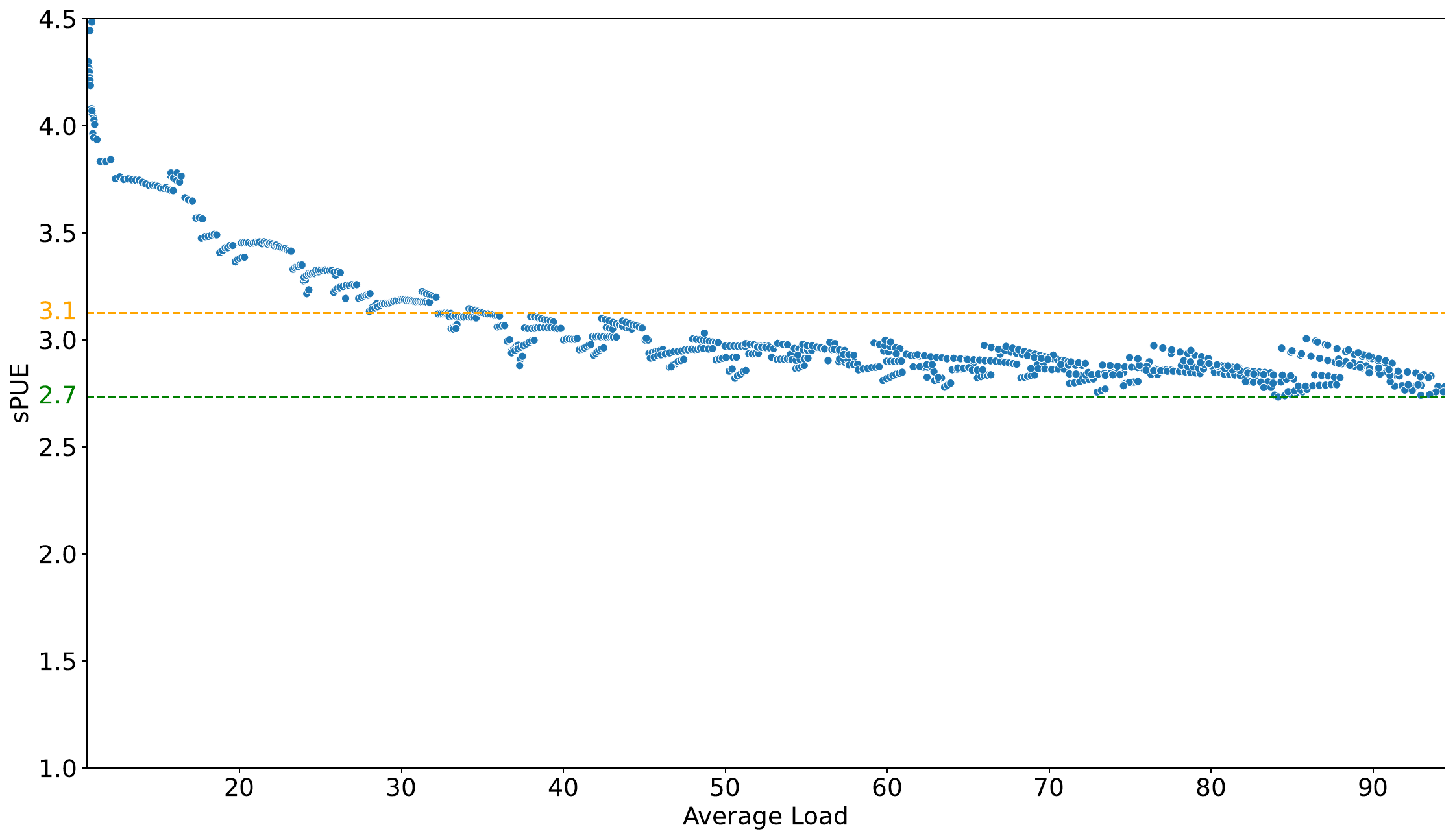}
	\caption{Correlation of the \gls{spue} and the CPU average load}
	\label{fig:xp:k8s:hpue-avgload}
\end{figure*}

Beyond this first observation, the \gls{spue} tends to decrease with higher CPU usage, no matter the number of concurrent containers, as acknowledged by Figure~\ref{fig:xp:k8s:hpue-avgload}.
With an average value of $3.1$ throughout the experimentation, one can observe that the optimal and lowest value reached for the \gls{spue} of such a standard server caps at $2.7$, which is high above the values commonly communicated for \gls{pue}.
As the \gls{spue} should be combined with the \gls{pue} of any cloud provider to report the \gls{gpue} reflecting the physical power consumption induced by any Joule consumed by the computing units (cf. Equation~\ref{eq:gpue}), it can only degrade the value of the standard \gls{pue}.
These observations thus challenge cloud infrastructures and \gls{maas} providers to deliver more energy-efficient servers by investing in energy-proportional techniques to power linearly a server with the triggered activities and computations.
This is particularly critical as \gls{dc} and cloud providers are known to be under-utilized~\cite{4404806,10.1145/3132747.3132772}, which let us expect higher \gls{spue} in production than the one we measured in our testbed under stress conditions.

\begin{figure}
    \centering
	\includegraphics[width=\linewidth]{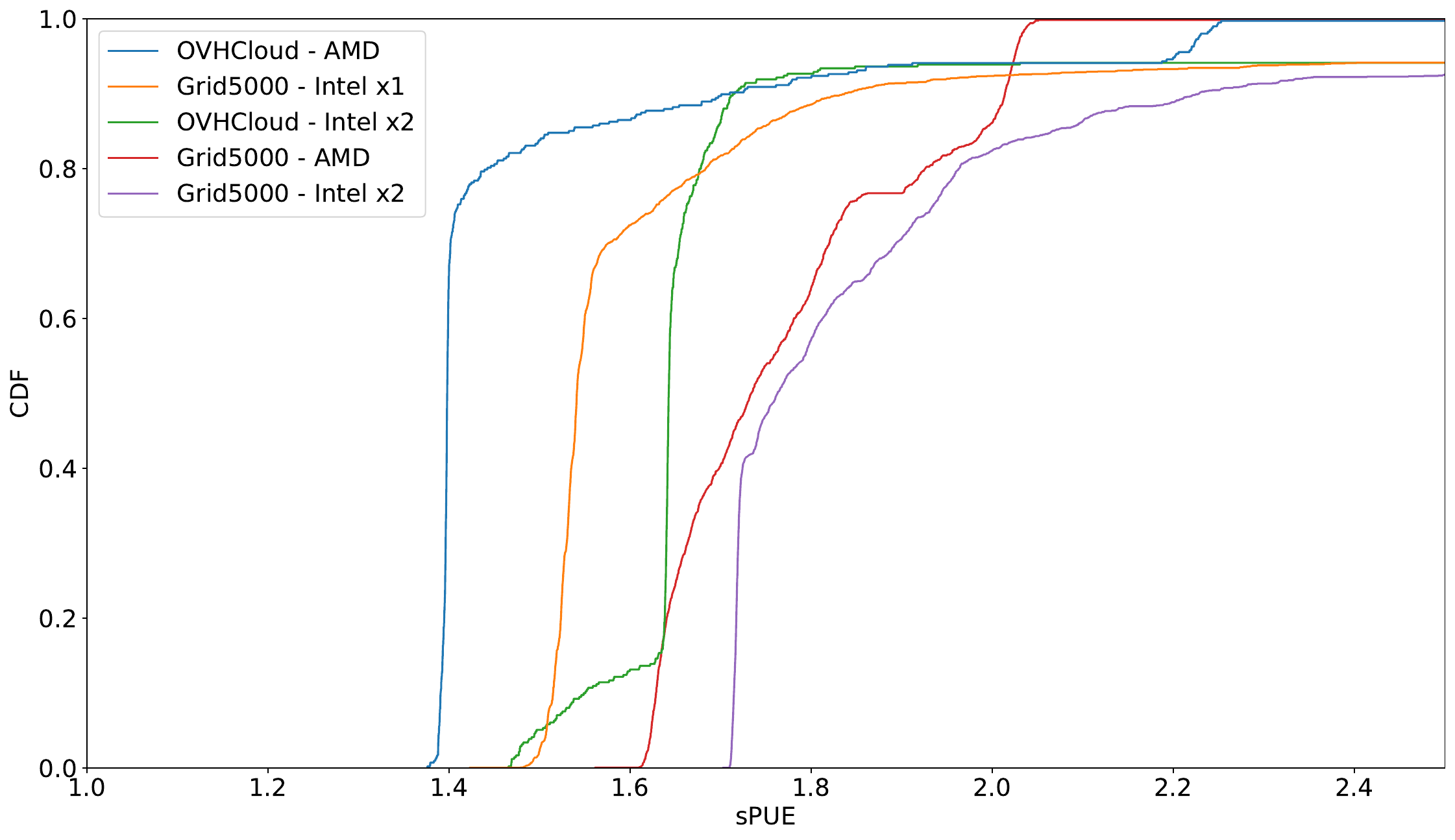}
	\caption{Comparing the \gls{spue} of all the hardware configurations under test.}
	\label{fig:xp:k8s:all-hpue}
\end{figure}

\paragraph{Hardware impact.}
To further investigate this hardware overhead of servers, we hypothesize like~\cite{barroso2013datacenter} that the \gls{spue} might be influenced by the CPU architecture and the cooling system of a server.
Therefore, using the same input workload as in the previous experiment, we compare the \gls{spue} of Intel and AMD servers that are deployed in Grid'5000 (based on air cooling) and \textsc{OVHcloud} (based on water cooling) infrastructures (cf. Table~\ref{table:validation:hardware-settings}).
While water cooling systems directly contribute to improving the \gls{pue} of the \gls{dc}, they may also indirectly impact the \gls{spue} by removing the CPU/GPU fans from the server frame, hence saving energy consumption induced by the equipment embedded in most of the servers.
Figure~\ref{fig:xp:k8s:all-hpue} thus depicts the \gls{cdf} of \gls{spue} for each hardware configuration of Table~\ref{table:validation:hardware-settings}.
Each configuration is stressed in the same conditions, by executing a workload of twice the maximum CPU load of each configuration to offer a fair comparison.
One can observe that the best \gls{spue} ($1.4$) results from the combination of a water cooling system and a single AMD CPU.
Interestingly, the optimal value for this configuration is quickly reached and maintained, compared to the other hardware configurations that report a wider distribution of \gls{spue} values.

\paragraph{Other impacts.}
While we could not change the AC-DC power supply of our testbed or include alternative supply designs, we believe that the \gls{spue} can also capture the efficiency of this hardware component and contribute to adopting a more energy-efficient solution.
We are also confident that this separation of concerns could be captured by a partial \gls{spue}, inspired by the \gls{ppue}~\cite{ISO:30134-2}, to isolate the overhead imposed by this hardware component.

One can therefore observe that optimizing the \gls{spue} does not only require adopting energy-saving strategies to power and cool down hardware components, but also maximizing the utilization of provisioned resources.
This observation thus challenges the hardware configuration of servers to be carefully sized to the closest number of CPU threads, and other hardware components, which are required to support a target workload.
In this context, elasticity mechanisms should be deployed at the hardware level by cloud infrastructure to implement energy proportionality and deal with the variability of workloads, hence preventing the over-provisioning of resources that require to be kept always on.

\begin{figure}
    \centering
	\includegraphics[width=.8\linewidth]{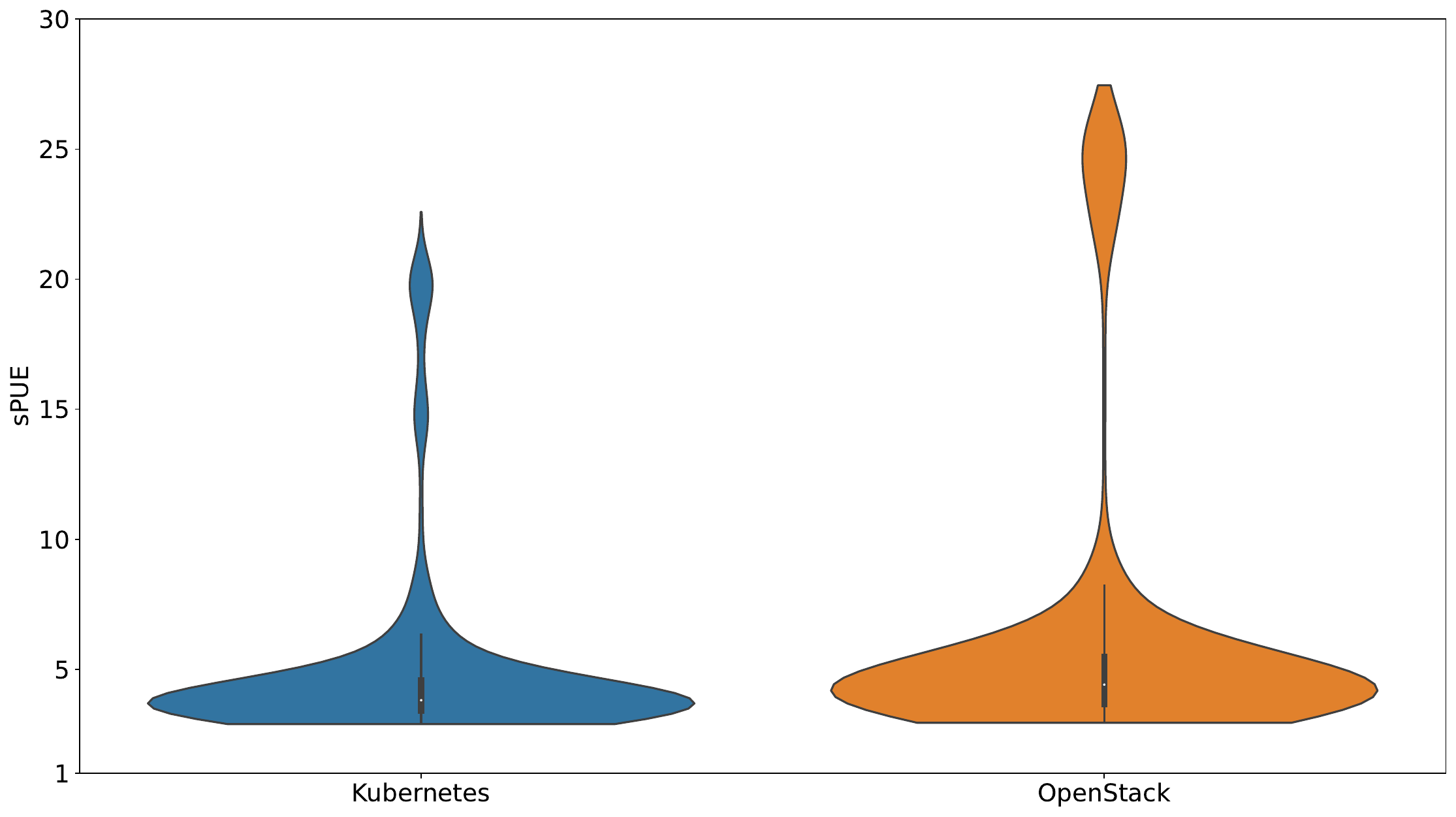}
	\caption{Comparing the \gls{spue} of a cluster used to host \textsc{OpenStack} and \textsc{Kubernetes} platforms.}
	\label{fig:xp:all:hpue}
\end{figure}

\paragraph{Cluster \gls{spue}.}
Beyond single node deployments, cloud infrastructures are often considered to provision a cluster of several nodes that are then assembled to deploy a \gls{iaas} (\emph{e.g.}, \textsc{OpenStack}) or \gls{caas} (\emph{e.g.}, \textsc{Kubernetes}) platform.
These clusters are typically composed of, at least, a control node and several worker nodes.
We, therefore, consider the injection of a cluster-wide input workload to both \textsc{OpenStack} and \textsc{Kubernetes} platforms, deployed atop 5 Intel Xeon Gold 5220 servers, to study the average \gls{spue} at the scale of a cluster, as mentioned in Section~\ref{sec:hpue}.
Table~\ref{table:cluster:hpue} and Figure~\ref{fig:xp:all:hpue} compare the \gls{spue} distribution of all the nodes involved in the cluster for both experiments.
Similarly to the case of idle servers, idle platforms may be the root cause of a critical \gls{spue} observed at the scale of a cluster, with observed factors above $20$.
Such situations typically witness the over-provisioning, and under-allocation, of a cloud infrastructure that allocates much more computing resources than required.
While the optimal \gls{spue} observed for a cluster of 5 nodes reaches $2.9$, one can observe that even a stressful scenario like one of our scenario results in an average \gls{spue} of $5$ to $6.5$---depending on platforms---thus highlighting the critical impact of the power usage efficiency of the nodes composing the cluster required to host a cloud infrastructure.

\begin{table}\small
    \centering
    \caption{Cluster-wide \gls{spue} statistics.}
    \label{table:cluster:hpue}
	\resizebox{.8\linewidth}{!}{%
    \begin{tabular}{l|c|c|c|c}
    \toprule
    \bf Platform        & \bf min & \bf max & \bf mean & \bf median \\
    \toprule
    \toprule
	\textsc{Kubernetes} & \color{orange}{$2.9$} & \color{red}{$22.6$} & \color{orange}{$5$}   &      $3.8$ \\
	\textsc{OpenStack}  & \color{orange}{$2.9$} & \color{red}{$27.5$} & \color{orange}{$6.5$} &      $4.4$ \\
    \bottomrule
    \end{tabular}
	}
\end{table}

\paragraph*{Network impact.}
In the context of a distributed setting, like a cluster, the question of including network equipment might arise.
As part of our experiment, we consider the impact of the network---\emph{i.e.}, a single \gls{tor} switch in our context---as part of this cluster setting, assuming that a switch is required to connect several nodes.\footnote{Production-scale cloud infrastructures may involve multiple \gls{tor} switches to ensure failover.}
However, the power consumption of most hardware network equipment (switch, routers, etc.) is known to be stable, no matter the workload ($115.3\,W$ in the context of our \gls{tor} switch), thus we advocate including the energy consumption of network equipment within the \textsf{IT} part of the \gls{spue}, and not as part of the \textsf{hardware} (cf. Equation~\ref{eq:hpue}).
The motivation for doing such is that \gls{spue} could be easily reduced by adding more and more network equipment in the \textsf{hardware} part (divisor), which would go against the objective of optimizing the energy efficiency of the overall cloud infrastructure.

\begin{table}\small
    \centering
    \caption{Control plane impact on \gls{spue}.}
    \label{table:control:hpue}
	\resizebox{.8\linewidth}{!}{%
    \begin{tabular}{c|c|c|c|c|c}
		\toprule
		\bf Node & \bf Platform        & \bf min & \bf max & \bf mean & \bf median \\
		\toprule
		\toprule
		control  & \textsc{Kubernetes} & \color{red}{$2.9$} & $4.4$ & \color{red}{$4$} &    $4$ \\
		node     & \textsc{OpenStack}  & \color{red}{$2$}   & $6.2$ & \color{orange}{$3.3$} &  $2.9$ \\
		\hline
		worker   & \textsc{Kubernetes} & \color{green}{$1.4$} &  $6.7$ & \color{orange}{$2.1$} &  $1.8$ \\
		nodes    & \textsc{OpenStack}  & \color{green}{$1.7$} & $14.9$ & \color{orange}{$3.5$} &  $2.8$ \\
		\bottomrule
	\end{tabular}
	}
\end{table}

\begin{figure}
	\centering
	\includegraphics[width=\linewidth]{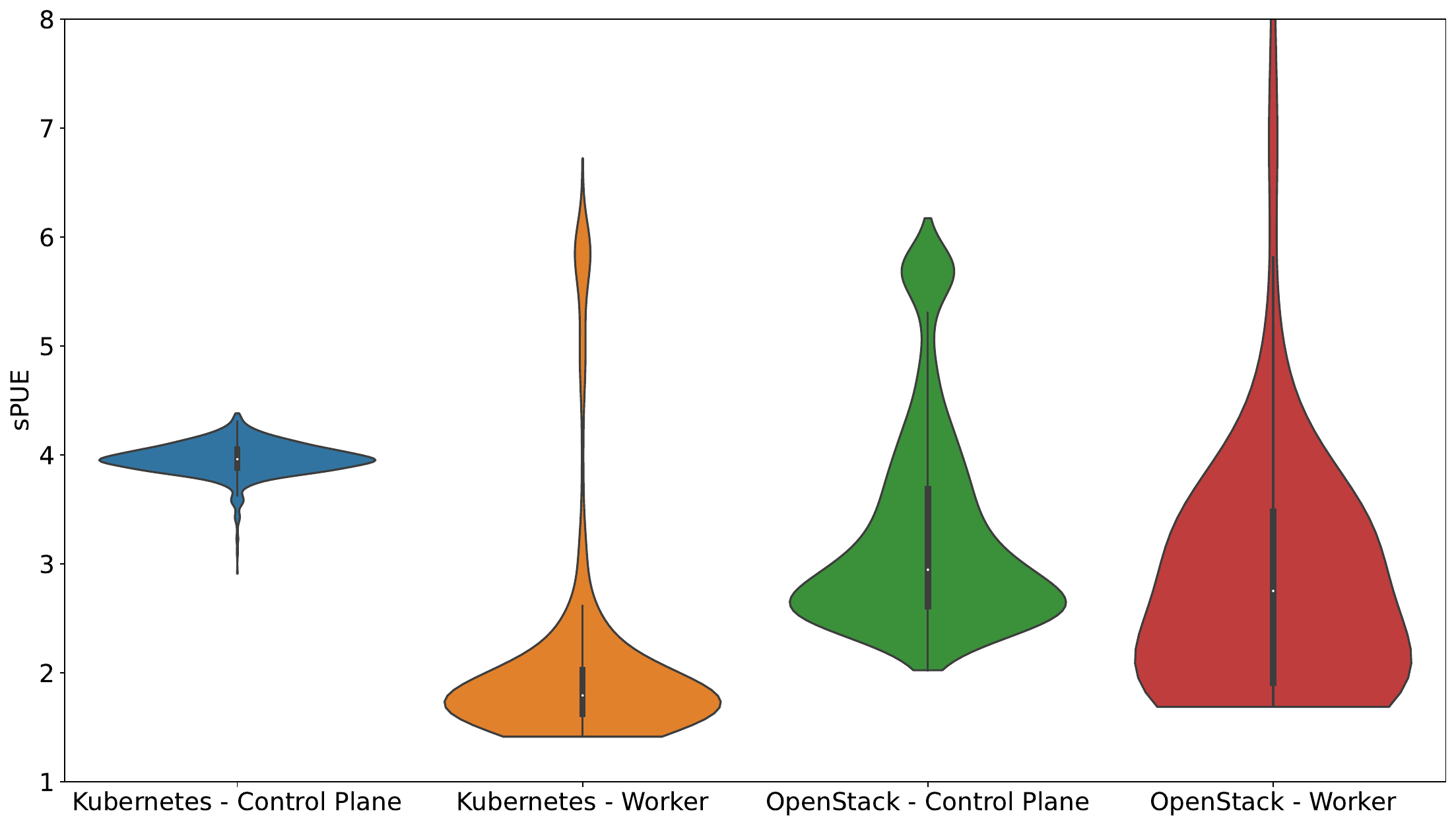}
	\caption{Comparing the \gls{spue} of control \& workers nodes for \textsc{OpenStack} \& \textsc{Kubernetes} platforms.}
	\label{fig:xp:all:hpue:control}
\end{figure}

\paragraph*{Control plane impact.}
Then, we compute the \gls{spue} of the control node and the worker nodes separately for each configuration (cf. Table~\ref{table:control:hpue} and Figure~\ref{fig:xp:all:hpue:control}).
One can observe that no matter the cloud platform, the \gls{spue} of the control node is always higher than the \gls{spue} of the worker nodes, as the control plane is consuming fewer resources than the worker nodes, on average, and thus, fails to exploit the full efficiency of the underlying hardware configuration.
One can nonetheless observe that an \textsc{OpenStack} control plane reports a slightly lower \gls{spue} on average, compared to its worker nodes.
This can be explained by the consumption of this control node, which tends to be relatively higher than other nodes, due to the number of control services that are involved in the \textsc{OpenStack} platform.
To reduce the impact of the control plane on the \gls{spue}, one should therefore consider maximizing the number of active worker nodes and consider the deployment of carefully sized control nodes, involving potentially smaller servers.

One can also observe that the \gls{spue} of \textsc{Kubernetes} and \textsc{OpenStack} platforms seem to differ, which highlights that both technologies are not stressing the hardware components in the same way.
To better understand the root cause of these differences, we further explore the \gls{vpue} of both platforms in the following section.

\subsection{\gls{vpue} Experiments}\label{sec:xp:vpue}
We then move to the study of \gls{vpue}, first in the context of an \gls{iaas} platform, based on \textsc{OpenStack}, and then on a \gls{caas} platform, based on \textsc{Kubernetes}.
Through the following experiments, we intend to study the resource overhead imposed by the software platform and all the services it provides to deploy and control \gls{vm} and/or containers.
We believe that such software layers represent another key efficiency factor to carefully consider when delivering cloud infrastructures.
Furthermore, unlike hardware layers, software layers can be embedded to deliver, for example, a \gls{caas} platform atop \gls{iaas}.
Thus, considering the impact of such common practices is also another insightful feedback that we intend to cover thanks to the \gls{vpue} indicator.

\subsubsection*{Estimating the \gls{vpue} of \textsc{OpenStack}}
\paragraph{Platform setting.}
To estimate the \gls{vpue} of \textsc{OpenStack}, we used the same input workload as in the \gls{spue} experiments, namely, we run a benchmark designed to start twice the maximum CPU load of the provisioned cluster.
This aims to investigate the efficiency of the cloud infrastructure ranging from an idle state to a situation of resource over-commitment.
We use a separate control node for hosting cluster-wide \textsc{OpenStack} services in addition to services that are required to be deployed in the worker nodes.
The CPU and DRAM overcommitment parameters are kept to default: \texttt{16:1} and \texttt{1.5:1}, respectively.
Each allocated \gls{vm} uses a profile \texttt{m1.exp}, which consists of 1 vCPU and 256\,MB of DRAM.
Given that we use 4 worker nodes, summing to $144$ threads and about $384$\,GB of DRAM, we can allocate---in theory---up to $2,304$\,vCPUs and $576$\,GB of vRAM, which represent about $2,000$ \gls{vm} with the profile \texttt{m1.exp}.
However, as previously mentioned, we stop our benchmark when reaching $288$ \gls{vm}, which represents twice the hardware threads made available at the scale of the cluster.

The \gls{vpue} is computed as the ratio of the total energy consumption of all the cluster services (including \gls{vm}) to the energy consumption of hosted \gls{vm}.
As introduced in Section~\ref{sec:powerapi}, we estimate the energy consumption of individual \gls{vm} by using the \textsc{SmartWatts} software-defined power meter~\cite{fieni:2020}.

\begin{figure}
    \centering
	\includegraphics[width=\linewidth]{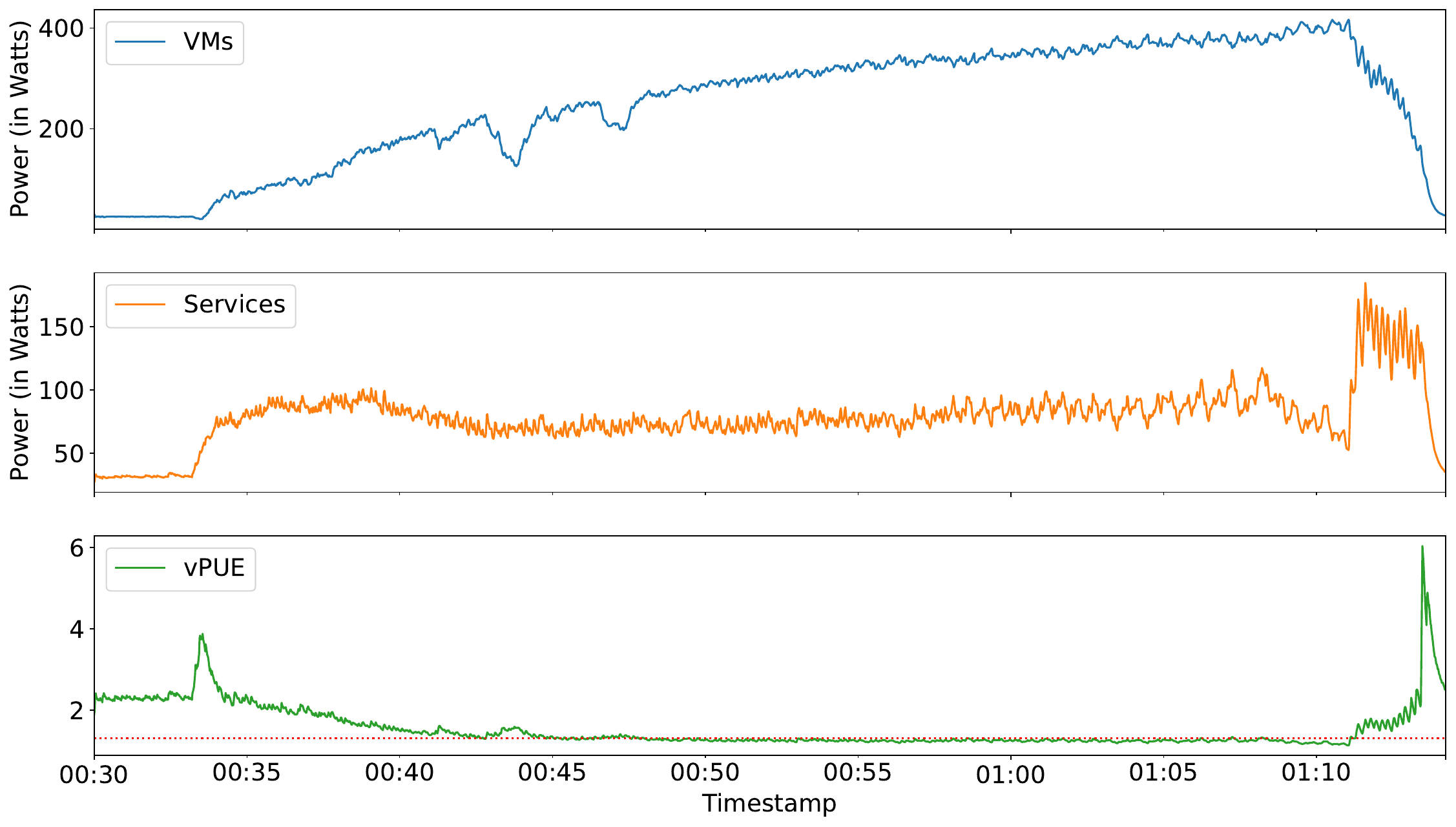}
	\caption{Evolution of the \gls{vpue} of \textsc{OpenStack} when increasing the number of hosted \gls{vm}.}
	\label{fig:validation:openstack:vpue}
\end{figure}

\paragraph{Overcommitment impact.}
Figure~\ref{fig:validation:openstack:vpue} illustrates the evolution of the \gls{vpue} over time when increasing the number of hosted \gls{vm}.
One can observe that the more \gls{vm}, the better \gls{vpue}, as one could expect.

\begin{figure}
    \centering
	\includegraphics[width=\linewidth]{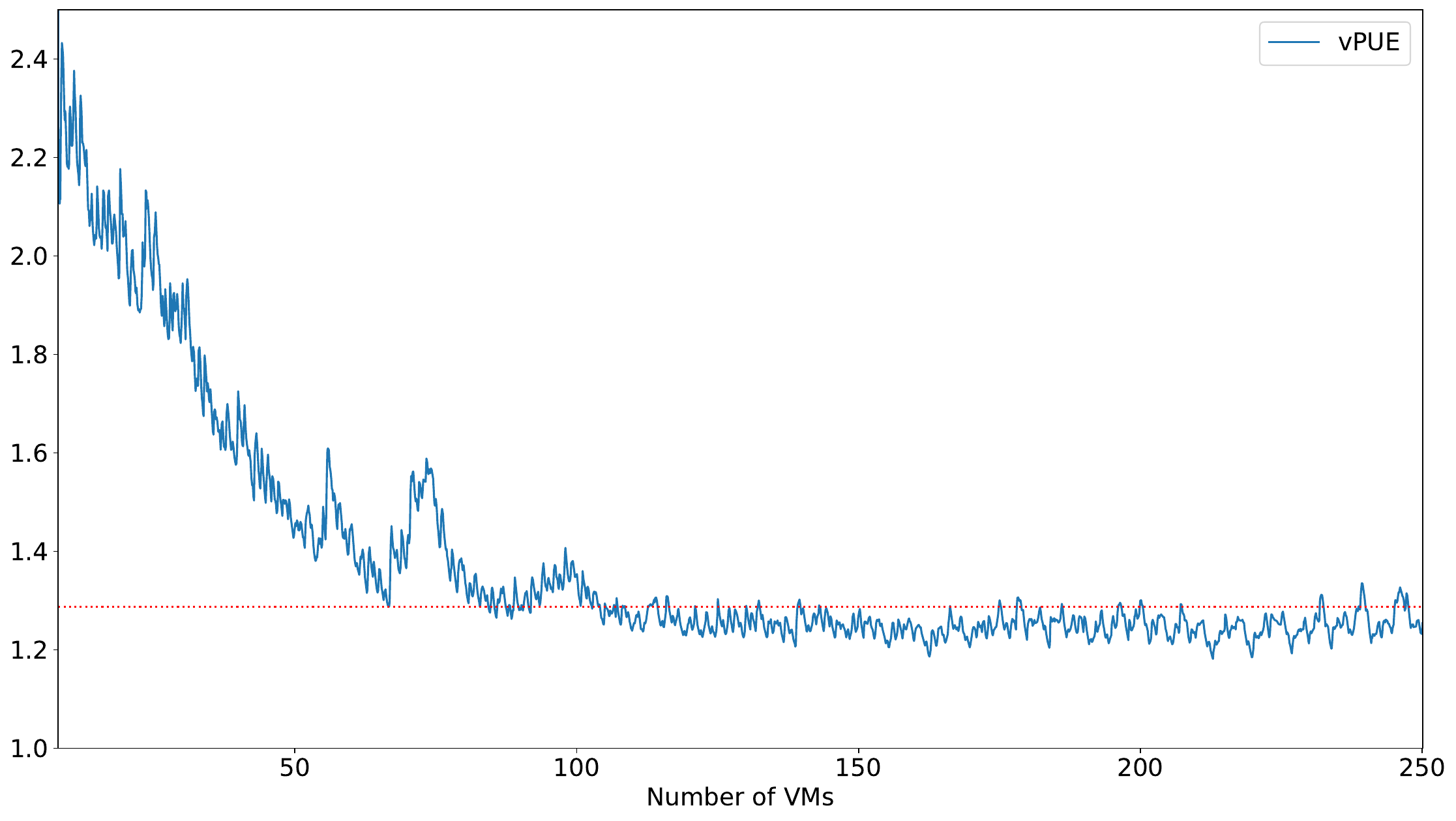}
	\caption{Correlation of the \gls{vpue} with the amount of provisioned \gls{vm}}
	\label{fig:validation:openstack:vpue-num-vms}
\end{figure}

More precisely, Figure~\ref{fig:validation:openstack:vpue-num-vms} shows that the \gls{vpue} converges towards its optimal value when reaching more than $100$ hosted \gls{vm}, which roughly corresponds to the total amount of physical resources (CPU \& memory) available in the cluster we provisioned.
This average value is estimated to $1.3$ in the context of our deployment of \textsc{OpenStack}, involving 1 control node and 4 worker nodes.
This first part of the experiment demonstrates that, as does the \gls{spue}, the optimal \gls{vpue} is reached when fully stressing the worker nodes, which encourages the optimization of the overcommitment parameters as a way to maximize the resource utilization of cloud infrastructures, hence favorably contributing to both indicators.

\paragraph{Control plane impact.}
Regarding the control plane of \textsc{OpenStack}, one can leverage the \textsc{SmartWatts} power measurements reported at the scale of individual services.
In particular, one can observe %
that the major power-consuming service is \textsf{neutron-api}, which is the central service for managing virtual machines in \textsc{OpenStack}.
In our testbed, \textsc{neutron-api} consumed $5\times$ more than the average power consumption of a hosted \gls{vm} %
, which can be partially explained by the benchmark we run to stress the \textsc{OpenStack} platform.

Yet, when considering the overall energy consumption for the scenario we executed, one can observe %
that almost 80\% of the energy is consumed by the virtual machines, and 20\% are imposed by the \textsc{OpenStack} services, resulting in an aggregated \gls{vpue} of $1.25$.

We believe that such detailed reports offered by partial \gls{vpue} and \textsc{SmartWatts} software-defined power meter can be further exploited by the operators of \textsc{OpenStack} platforms to highlight platform energy hotspots that require to be carefully considered, hence challenging the relevance and benefit of deployed control services, beyond the standard configurations. 

One should also note that, while the \gls{vpue} reports a lower value ($1.25$) than the \gls{spue} ($2.7$, in Figure~\ref{fig:xp:all:hpue}), it keeps representing a factor that needs to be combined with other indicators---\emph{i.e,} $\gls{cpue} = 2.7 \times 1.25 = 3.375$ in this experiment---to reason upon global indicators and not partial ones.
This \gls{cpue} roughly indicates that $1$\,Watt consumed by any hosted \gls{vm} effectively imposes $3.375$\,Watts at the scale of the cluster, in the best case.

\subsubsection*{Estimating the \gls{vpue} of \textsc{Kubernetes}}
Beyond the specific case of \textsc{OpenStack}, we also investigate the \gls{vpue} of a \textsc{Kubernetes} platform in this section.

\paragraph{Platform setting.}
We keep using our benchmark designed to start twice the maximum CPU load of the cluster.
To keep the configuration of the \textsc{Kubernetes} cluster as close as possible to a production environment, we follow the best practices and do not allow the scheduling of containers to the control plane.
This means that, in a cluster composed of $4$ worker nodes, there are $144$ available threads.
During this experimentation, we observe an average \gls{vpue} of $1.3$ with an optimal value of $1.1$.
Interestingly, one can observe that the control plane of \textsc{Kubernetes} consumes less energy than the one of \textsc{OpenStack} to execute the same workload.
One can observe in Table~\ref{table:cluster:consumption} that the services composing the control plane of \textsc{OpenStack} impose an overhead of $151.1~kJ (910\%)$ compared to the control plane of \textsc{Kubernetes}, while the expectable overhead imposed by virtual machines over containers is limited to $109.6~kJ (23\%)$.
The control plane of \textsc{Kubernetes} thus leaves $96.88\%$ of the total energy consumed for the execution of hosted containers, compared to $78.08\%$ in the case of \textsc{OpenStack}.

\begin{table}\small
    \centering
    \caption{Energy consumption of \textsc{Kubernetes} and \textsc{OpenStack} services.}
    \label{table:cluster:consumption}
	\resizebox{.8\linewidth}{!}{%
    \begin{tabular}{c|c|r|r}
		\toprule
		\bf Node & \bf Platform        & \bf Energy\,(kJ) & \bf diff\,(kJ) \\
		\toprule
		\toprule
		control  & \textsc{Kubernetes} & $15.6$ & \\
		services & \textsc{OpenStack}  & \color{red}{$\textbf{166.7}$} & \color{red}{$\textbf{+151.1}$} \\
		\hline
		hosted & \textsc{Kubernetes} & $484.3$ & \\
		jobs   & \textsc{OpenStack}  & $\textbf{593.9}$ & $\textbf{+109.6}$\\
		\bottomrule
	\end{tabular}
	}
\end{table}

\subsection{\gls{cpue} \& \gls{gpue} Experiments}\label{sec:xp:cpue}
As previously mentioned, \textsc{Kubernetes} clusters can be provisioned atop a set of \gls{vm} hosted by a \gls{iaas}.
In such a situation, the \gls{vpue} of the \gls{iaas} has to be combined with the one of \textsc{Kubernetes} to better reflect the resulting efficiency of the platform.

\begin{figure}
    \centering
	\includegraphics[width=\linewidth]{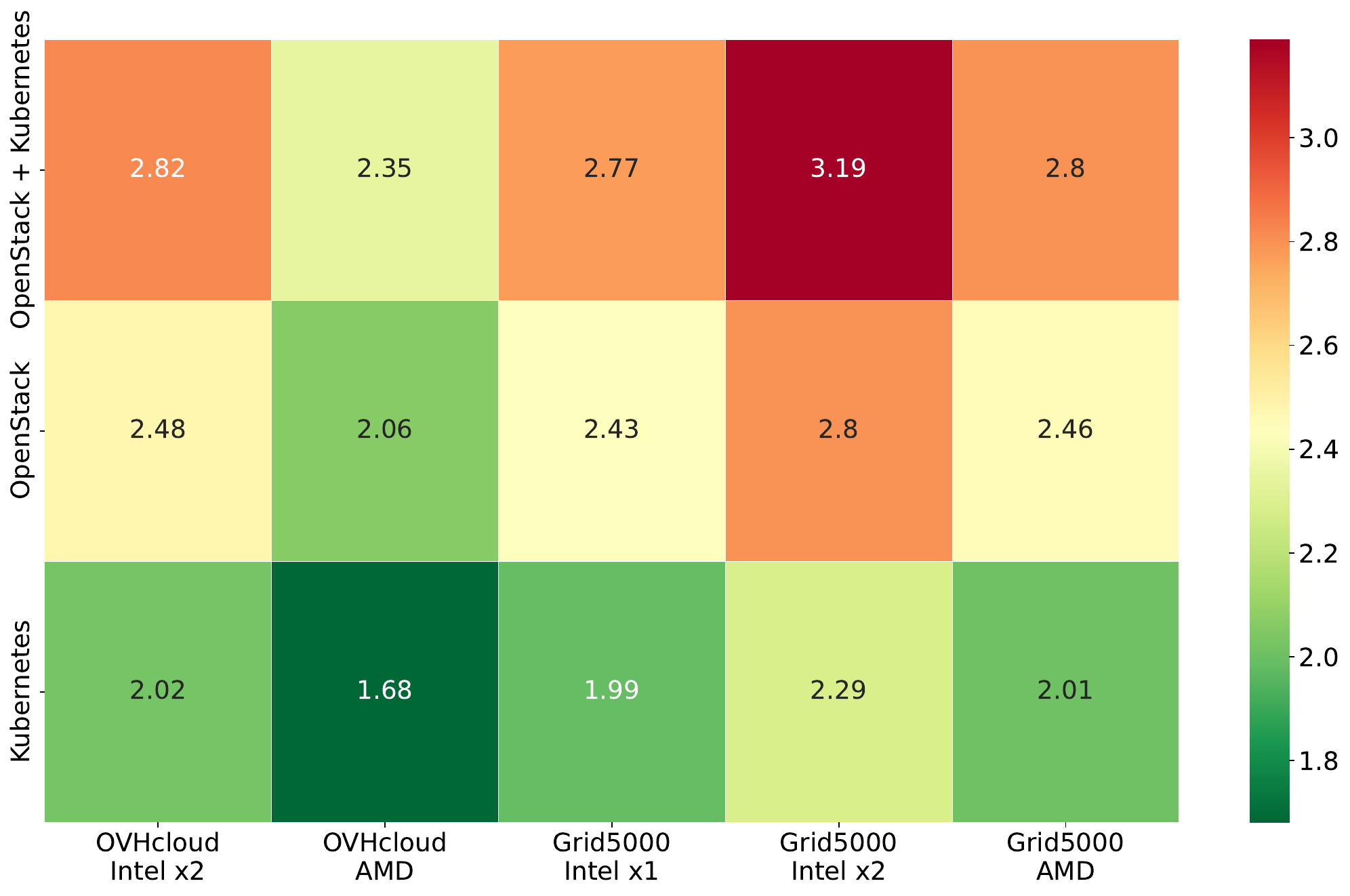}
	\caption{Comparing the \gls{cpue} of hardware/software configurations.}	
	\label{fig:validation:cpue-per-configuration}
\end{figure}

To evaluate the ability of \SYS{} to assess the energy efficiency of a \textsc{Kubernetes} infrastructure provisioned in an \textsc{OpenStack} cluster, we explore different configurations of \gls{iaas}/\gls{caas} technologies and compute the associated \gls{cpue}.
The estimated \gls{cpue} metrics are summarized in Figure~\ref{fig:validation:cpue-per-configuration}.
By combining \gls{spue} and \gls{vpue} metrics, one can observe that both the hardware server and the software stack can have a strong influence on the energy consumption of cloud infrastructures.
In our setup, at the scale of a single server, one can observe that the \gls{cpue} can range from $1.68$ for a baremetal \textsc{Kubernetes} cluster hosted by AMD servers provisioned by \textsc{OVHcloud} to $3.19$ for a virtualized \textsc{Kubernetes} cluster provisioned atop a set of \textsc{OpenStack} virtual machines deployed in a private \gls{iaas} (Grid5000).
This observation strengthens our claim that the optimization of energy efficiency of cloud infrastructures requires a holistic approach covering all hardware and software layers, beyond the sole optimization of the \acrlong{dc} and its \gls{pue}.

\begin{figure}
    \centering
	\includegraphics[width=\linewidth]{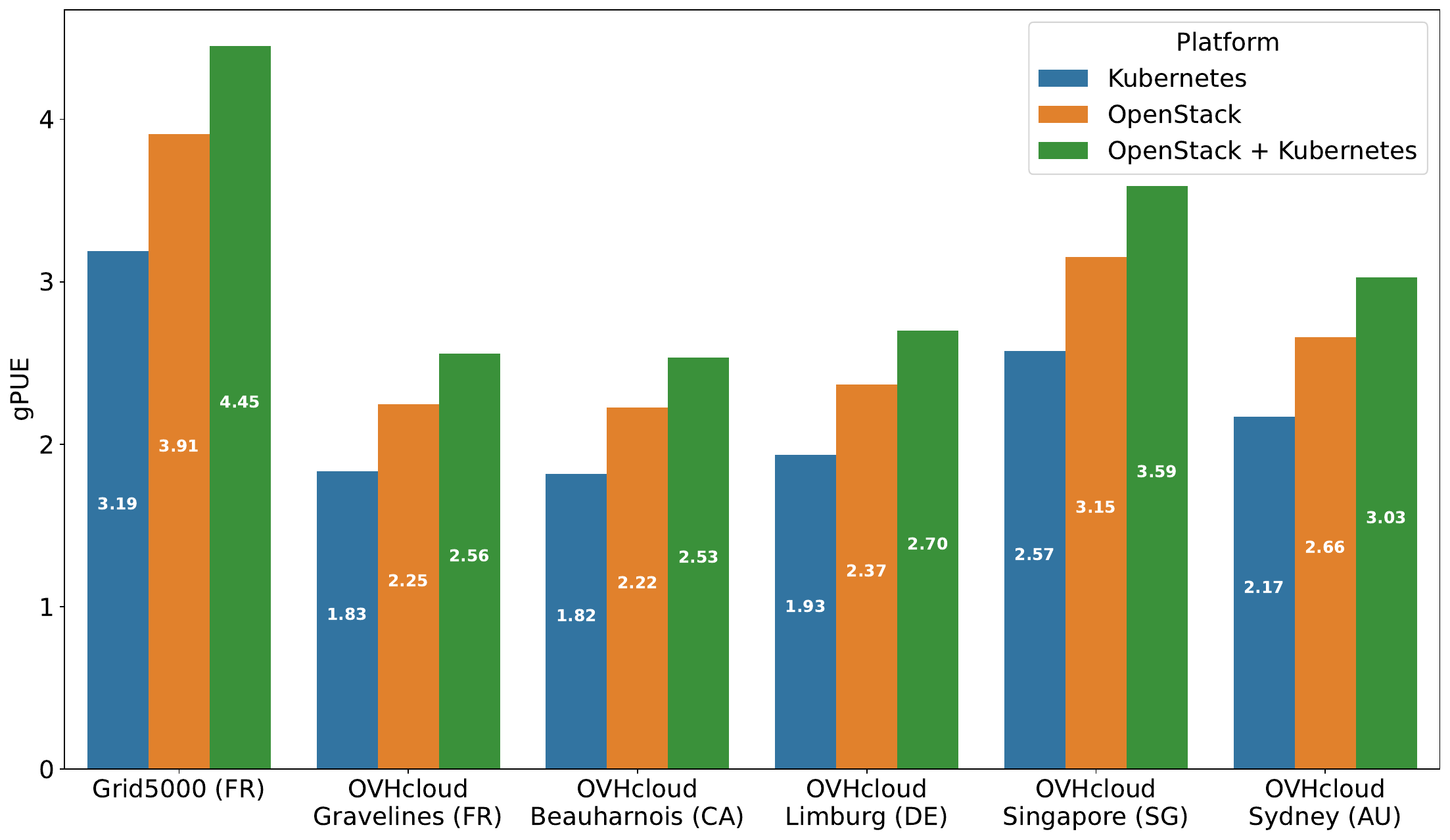}
	\caption{Comparing the \gls{gpue} of \gls{dc}/platform configurations.}
	\label{fig:validation:gpue-per-dc}
\end{figure}

To further reveal the concrete energy footprint of a cloud service, one should therefore combine the \gls{cpue} of the cloud infrastructure with the \gls{pue} of the data center hosting the servers.
Figure~\ref{fig:validation:gpue-per-dc} estimates \gls{gpue} of different deployments of \textsc{OpenStack} and \textsc{Kubernetes} hosted on an AMD server provisioned by \textsc{OVHcloud}, according to the selected \gls{dc}.				
We use the \gls{pue} publicly published as a reference for the \gls{dc} built by \textsc{OVHcloud}---\emph{i.e.}, Gravelines in France, Beauharnois in Canada, and Limburg in Germany.
As for the \gls{dc} leased to \textsc{OVHcloud}---\emph{i.e.} \href{https://www.bdxworld.com/}{Big Data Exchange} (previously Telstra) for Singapore and \href{https://www.nextdc.com/}{NextDC} for Sydney in Australia---we use the data publicly published by the operators as a reference.

One can observe that all the \glspl{dc} owned by \textsc{OVHcloud} report on a similar \gls{gpue}, no matter the operated cloud platform, which is more efficient than traditional \glspl{dc}.
The \gls{gpue} reported in Figure~\ref{fig:validation:gpue-per-dc} highlights the limitations of exposing the \gls{pue} in isolation, with values that are all above $1.8$, while the \gls{pue} is claimed to stagnate at $1.5$, according to the Uptime Institute~\cite{uptime-insitute:industry-pue-2023}.

\begin{figure}
    \centering
	\includegraphics[width=\linewidth]{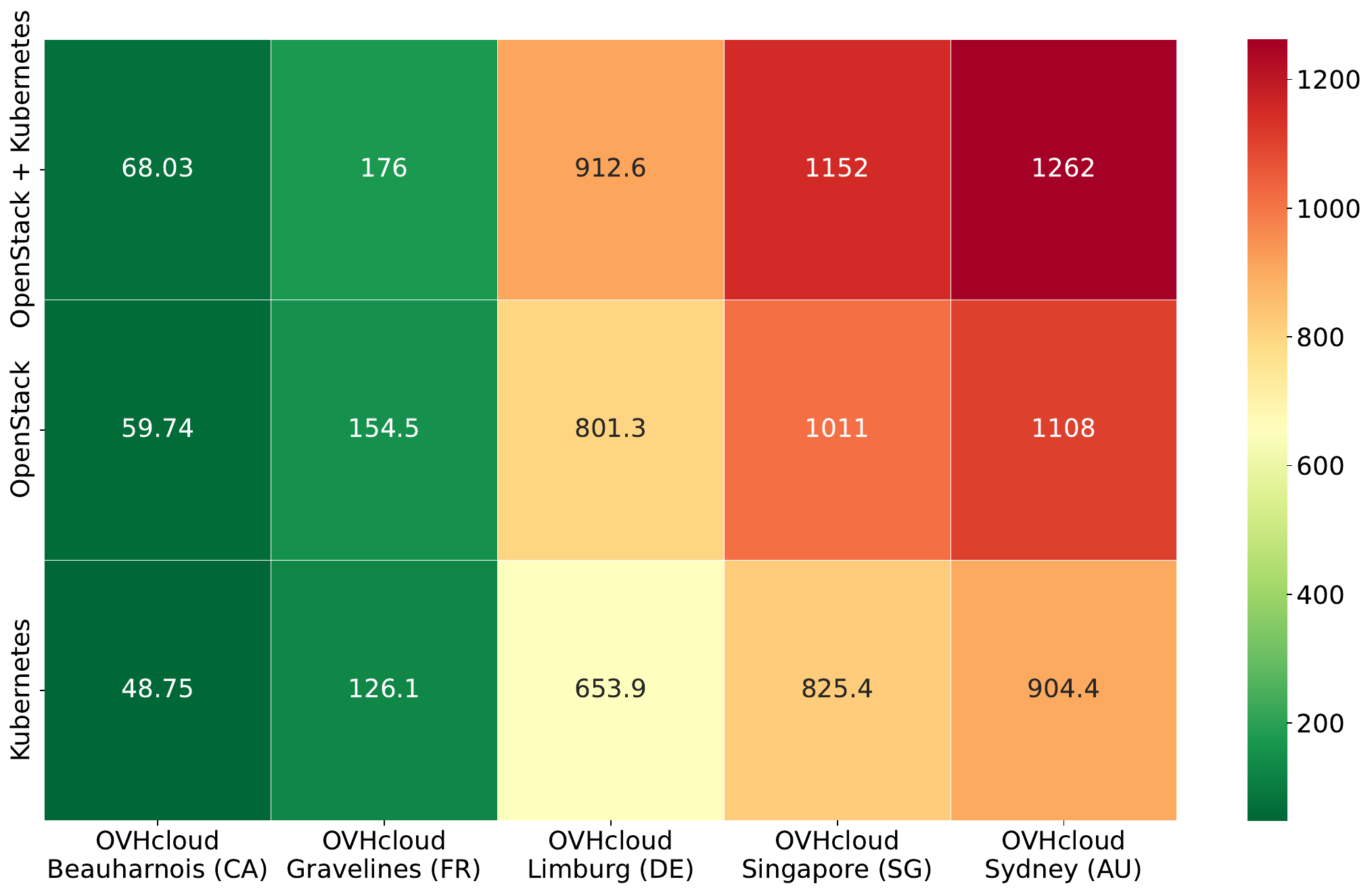}
	\caption{Comparing the \gls{gcue} of \textsc{OVHcloud} \gls{dc} of various countries.}
	\label{fig:validation:gcue-per-dc}
\end{figure}

Given the worldwide presence of \textsc{OVHcloud} and other cloud providers, we are also studying the impact of the effective \gls{dc} location.
This location has a direct impact on the energy mix adopted to power the \gls{dc}, which can be considered to estimate the related carbon emissions. 
Therefore, Figure~\ref{fig:validation:gcue-per-dc} estimates the \gls{gcue} of several \glspl{dc} of \textsc{OVHcloud} across the world.
As there is no data officially published by \textsc{OVHcloud} at the \gls{dc} granularity for the \gls{cue}, we choose to rely on the data published by \href{https://www.electricitymaps.com/}{\textsc{Electricity Map}} as a reference for the \emph{Carbon dioxide Emission Factor} (CEF) of the countries where are based each \gls{dc}.

By extending the \gls{cue} with the \gls{cpue} we described in this article, we observe that the range of \gls{gcue} is further extended, with $48.75$\,eqCO2 per unit of computation for operating software in a \textsc{Kubernetes} cluster to $1,262$\,eqCO2 for the same unit of computation operated in Sydney in a \textsc{Kubernetes} cluster that is deployed atop \textsc{OpenStack}.
This figure shows that the combined choice of \gls{dc} location, server configuration and software stacks can have a strong impact on carbon emissions, as illustrated by the factor $26$ observed between the two extreme values of Figure~\ref{fig:validation:gcue-per-dc}.

We believe that the adoption of \SYS{} metrics can not only help cloud providers to better understand the impact of their choices on the energy and carbon footprint of their services but also help their customers to make more informed decisions when selecting a cloud provider.
We also hypothesize that \SYS{} metrics can stimulate the hardware and software industries to design and implement more energy-efficient solutions, which can be further validated by \SYS{} metrics.

%% file: sections/recommendations.tex
\section{Recommendations}\label{sec:recommendations}
Based on the above experiments, we can formulate the following recommendations for the design of energy efficient cloud platforms. 
These recommendations are targeting both cloud providers and cloud users, depending on the level of control they have over the cloud infrastructure:
\begin{itemize}
    \item \textbf{Impact of cooling}. 
    We observed that liquid-cooled servers report better energy efficiency than traditional air-cooled servers (cf. Figure~\ref{fig:xp:k8s:all-hpue}). 
    Liquid-cooling is one example of passive cooling techniques, which can dissipate more heat per watt, compared to active techniques.
    We, therefore, recommend to prioritize the provisioning of passively-cooled servers operated by cloud operators.
    
    \item \textbf{Impact of hardware}. 
    Although they exhibit higher TDP properties (cf. Table~\ref{table:validation:hardware-settings}), AMD processors reported better energy efficiency than Intel processors, on average. 
    As observed in Figure~\ref{fig:xp:k8s:all-hpue}, reported numbers can be influenced by the model of processors, its generation, its chip technology.
    Beyond the impact of the cooling system, the chassis  can also influence the energy efficiency of the servers.
    Given this combination of factors, we recommend to report \emph{partial} \gls{spue} of individual hardware components, such as CPU, GPU, as well as the cooling system, the power supply and the chassis in order to encourage the provisioning of energy-efficient hardware components.

    \item \textbf{Impact of usage}. 
    Server usage should be maximized to optimmize the energy efficiency. This can be done by consolidating workloads on fewer servers~\cite{DBLP:conf/ic2e/HavetSFCRF17}, turning off idle \glspl{vm} and servers~\cite{DBLP:conf/ic2e/ZhangARPS17}, and using power management features to reduce power consumption. 
    Cluster-wide oversubscription~\cite{DBLP:conf/cluster/JacquetLR24}, as well as dynamic~\cite{DBLP:journals/tsusc/JacquetLR24} and vertical~\cite{DBLP:conf/ccgrid/JacquetLR24} oversubscription are alternative techniques to increase the utilization rate of cloud servers and \glspl{vm}, while preserving their quality of service.

    \item \textbf{Impact of clusters}.
    We observed that the control plane may induce a non-negligible impact on the energy efficiency of a cloud platform (cf. Figure~\ref{fig:xp:all:hpue:control}): the more worker node, the lower impact of the control plane.
    We, therefore, advocate considering the adoption of large clusters in cloud data centers, possibly subscribing public/managed cloud offers, instead of private cloud deployments.
    Given this observation, and the above-reported impacts, increasing the size of any cluster can only be beneficial if the usage of individual servers is maximized.
    Additionally, the hardware configuration of control plane servers can be optimized to reduce their energy consumption, as well as the number of control plane servers, according to our observations.

    \item \textbf{Impact of virtualization}.
    We observed that software containers are reporting better energy efficiency than \glspl{vm}.
    Nonetheless, the energy efficiency of \glspl{vm} remains competitive, as long as it is not considered as a means to implement nested virtualization, like hosting a container runtime inside a \gls{vm} (cf. Figure~\ref{fig:validation:cpue-per-configuration}).
    When possible, we recommend the deployment of containers in public cloud infrastructures.

    \item \textbf{Impact of energy mix}.
    Furthermore, when going beyond energy efficiency, and considering carbon emissions of cloud infrastructures, we observed that the physical location of \glspl{dc} has a critical impact on carbon emissions (cf. Figure~\ref{fig:validation:gcue-per-dc}).
    Our end-to-end \SYS metrics highlight that the choice of physical location is of utmost importance due to the amplification the emissions imposed by the combination of energy-efficiency factors. %

    \item \textbf{Impact on water consumption}.
    As mentioned in Section~\ref{sec:gmetrics}, one can observe that nothing prevent to combine the \gls{cpue} compound metric we propose (cf. Section~\ref{sec:cpue}) with the state-of-the-art \gls{wue} metric in order to compute the \gls{gwue} indicator, which can be used to assess the water consumption of cloud infrastructures as follows:
    \begin{equation}\label{eq:gwue}\scriptsize
        \gls{gwue} = \gls{cpue} \times \gls{wue} = \frac{\gls{cpue} \times \sum{AnnualWater(\textsf{DC})}}{\sum{Energy(\textsf{IT})}}
    \end{equation}
    By optimizing their \gls{cpue}, cloud stakeholders can effectively assess the water consumption of cloud infrastructures with the resulting indicator, \gls{gwue}, and compare the water consumption of different cloud infrastructures, which is another environmental \gls{kpi} that is increasingly adopted by cloud providers.

\end{itemize}

%% file: sections/conclusion.tex
\section{Conclusion}\label{sec:conclusion}
The energy efficiency of cloud infrastructures is a critical concern for service deployments and a lot of work has been made to accurately evaluate this efficiency.
While multiple indicators aim to assess the efficiency of data centers, none of them takes into account the efficiency of the hosted software.
They mostly treat the hosted software as a black box and report yearly feedback about the global efficiency of the infrastructure, which does not allow the evaluation of specific parts of the infrastructure. 

To address this issue, we propose \SYS{}, a family of \acrlong{pue} metrics that deliver a complementary perspective on the energy efficiency of cloud infrastructures, which remain uncovered by the state of the art.
In this paper, we report on the implementation of \SYS{} metrics using the \powerapi{} framework and its \textsc{SmartWatts} power meter to assess the impact of all the layers of a cloud infrastructure, from hardware settings to the software platforms.
While we demonstrate our approach on \textsc{Kubernetes} and \textsc{OpenStack} test clusters, we strongly believe that it can be deployed and used across a large variety of infrastructures. 
We took care to allow high flexibility to the users to easily adapt our solution to their specific infrastructure control plane and tools.

Furthermore, thanks to \SYS{}, cloud infrastructure operators can assess in real-time the efficiency of their infrastructure down to the software level.
This allows for more experiments on energy management policies, and faster feedback about their efficiency for the cloud operator.
Instead of relying on metrics that impose a long delay (mostly yearly) before getting feedback, we believe that \SYS{} provides a reliable indicator of the energy efficiency of the different parts of cloud infrastructure, while allowing more flexibility for the cloud operator.
In particular, we believe that cloud software architectures, like microservices, can leverage these online indicators to better guide the optimization of the resource consumption~\cite{DBLP:conf/sc/HouL0Z0G20,DBLP:conf/ipps/HouL0ZRLCG21} and influence the deployment and resources.

Beyond the optimization of the infrastructure, we believe that the end-to-end estimation of energy/carbon impacts of cloud software paves the way to reporting and tracking the environmental emissions of cloud software.
This is a critical step towards more sustainable cloud computing by raising awareness of customers on the hidden impact of their online activities.

\section*{Acknowledgments}
This work is supported by the ``{\em FrugalCloud}'' Inria and {\sc OVHcloud} partnership.
Additionally, this work also received partial support from the French government through the {\em Agence Nationale de la Recherche} (ANR) under the France\,2030 program, including partial funding from the {\sc CARECloud}~({\sf ANR-23-PECL-0003}), DISTILLER~({\sf ANR-21-CE25-0022}), and {\sc GreenAct} ({\sf ANR-21-HDF1-0006}) grants.

The maintenance of the PowerAPI toolkit is currently funded by Inria, Orange\,Labs, {\sc OVHcloud} and {\sc La\,Poste} Group.